\documentclass{aa}
\usepackage{txfonts}   %
\usepackage[dvips]{graphicx}
\usepackage[dvips]{epsfig}
\usepackage{url}
\usepackage{rotate}
\usepackage{rotating}
\usepackage{natbib}
\bibpunct{(}{)}{;}{a}{}{,} %
\def\3{\ss}                                                         
\def\cm{\vspace{1cm}}

\def\q0{\phantom{1}}

\def\al{\alpha}                                                         
                                                         
\def\si{\sigma}
\def\sig{\sigma}

\def\fe0{\frac {e}{E_{0}} }                                             
\def\eV0{\frac {e{V_{0}}}{E_{0}} }

\def\part{\frac{\partial}{\partial t}}

\def\m21{$2^{\circ}\times 1^{\circ}$}
\def\ts{\thinspace}

\def\oiv{O\ts{$\scriptstyle {\rm IV}$} }

\def\siv{Si\ts{$\scriptstyle {\rm IV}$}}
\def\ne8{Ne\ts{$\scriptstyle {\rm VIII}$} }
\def\fts{\footnotesize}
\def\arcsec{{\fts$^{\prime\prime}$}}

\begin{document}

\title{A nanoflare model of quiet Sun EUV emission}

\cm

\author{A. Pauluhn \inst{1,2} \and S.K. Solanki \inst{1}}

\institute{Max-Planck-Institut f\"ur Sonnensystemforschung, 
    D-37191 Katlenburg-Lindau, Germany  
\and 
now at Paul Scherrer Institut, CH-5232 Villigen, Switzerland}

\offprints{A. Pauluhn,
       \email{anuschka.pauluhn@psi.ch}}

\date{Received / Accepted }

\abstract{
Nanoflares have been proposed as the main source of heating of the solar 
corona. However, detecting them directly has so far proved elusive, and 
extrapolating to them from the properties of larger brightenings 
gives unreliable estimates of the power-law exponent $\alpha$ characterising their 
distribution. 
Here we take the approach of statistically 
modelling  light curves representative of the quiet Sun as seen in 
EUV radiation. 
The basic assumption is that all quiet-Sun EUV emission is due to 
micro- and nanoflares, whose radiative energies display a power-law 
distribution. 
Radiance values in the quiet Sun follow a lognormal distribution. 
This is irrespective of whether the distribution is made over a spatial 
scan or over  a time series. 
We show that these distributions can be reproduced by our simple model.  
By simultaneously fitting the radiance distribution function and  the power 
spectrum obtained from the light curves emitted by transition region 
and coronal lines
the power-law distribution of micro- and nanoflare brightenings 
is constrained. 
A good statistical match to the measurements is obtained 
 for a  steep power-law distribution of nanoflare energies, 
with power-law exponent $\al > 2$. This is consistent with the 
dominant heat input to the corona being provided by nanoflares, i.e., 
 by events with energies around $10^{23}$~erg.  
In order to reproduce the observed SUMER time series approximately 
$10^{3}$ to $10^{4}$ nanoflares are needed per second throughout 
the atmosphere of the quiet Sun (assuming the nanoflares to cover an 
average area of  $10^{13}$~m$^{2}$). 
}

\titlerunning{A nanoflare model of quiet Sun EUV emission} 
\authorrunning{Pauluhn \& Solanki}
\maketitle

\section{Introduction}

The solar emission in the EUV and X-ray wavelength range features transient 
events 
on all scales, such as flares 
and micro- or nanoflares, 
explosive events, and blinkers, see, e.g., \cite{tandberg} or 
\cite{solanki02} and references therein. 
Flares, micro- and nanoflares refer to similar phenomena with different 
magnitudes of energy release, which are usually studied in coronal emission, 
while explosive events and blinkers are spectral line broadenings and 
brightenings that are mainly restricted to the transition region. 
Events in the energy range $10^{30} - 10^{33}$~erg ($10^{23} - 10^{26}$~J) 
are usually referred to as ``normal'' flares. 
Micro- and nanoflares \citep{parker88} 
are  the brightenings 
with energy below approximately $10^{27}$~erg, although the limits vary 
somewhat in the literature.
Such transient brightenings are thought to 
be the signature of dissipation of the energy stored in the magnetic field,   
e.g., through magnetic reconnection. Thus,  they 
provide a mechanism for heating the solar corona 
\citep{parker88}. 

The frequency or rate distribution of the energy of flares 
($dN/dE$)  
has been found to obey a power law for several wavelength regimes.  
Power laws have also been found for blinkers by \cite{alen01} and for  
explosive events by \cite{winebarger02}.  
The distributions found by  \cite{winebarger02}, however, look more consistent 
with a lognormal function than a power law. 
Flares and their distributions have been the subject of a large 
number of studies, e.g., \cite{datloweetal74, linetal84,krubenz98, parnelljupp00, Aschw00,urmila01, Aschpar02, 
veksteinjain}. 
Power laws have also been  applied for stellar flare energy distributions, 
see, e.g., \cite{audard00, guedel03,arznerguedel} and references therein.   
\cite{luham} have explained 
the power-law dependence of the solar flare occurrence rate in a model of 
self-organized criticality as avalanches of many small reconnection events. 
\\  
The observed exponents $\alpha$ in the power-law relation 
$dN/dE = E_{0} E^{-\alpha}$ range from $1.5$ to $2.9$ for solar and 
 for stellar flares, but most studies give values of $\al < 2$.   
Exponents within this range are also obtained 
for blinkers and explosive events. 
The wide range of the exponent is to a large part 
due to different selection 
criteria or assumptions about geometry and structure of the emitting plasma as 
well as instrumental resolution.  
The larger the exponent, the more weight is given to small-scale events 
such as micro- and nanoflares. 
For an exponent greater than 2, the energy content is dominated by 
the small-scale events, and in order to have finite total energy content, a 
lower cutoff in energy has to be introduced. 
Since the energy released in flares and flare-like events of sufficient 
size to be directly detected is not enough to explain the temperature of the 
corona, an exponent above 2 is necessary if a mechanism producing flaring is 
to be the main cause of coronal heating.  
Thus, the distribution of the smallest flares is of great interest in order 
to have reliable estimates on their participation in coronal heating. 

Most previous work has concentrated on identifying individual events in 
radiance time series. Clearly, it is easy to miss or underestimate the number 
of the weakest such events, which may lead to a background of almost 
continuous brightening if they are sufficiently common. 
Note that in order to judge the amount of energy released through transient events, the 
X-ray, EUV and  
UV emission is often decomposed into a (nearly) steady background 
of emission with superposed transient brightenings.  

In this work we start from the basic assumption that all the emission at EUV 
wavelengths is produced by transient events (i.e., the observed apparent background 
is a superposition of many such events). 
We model in a simple way 
time series of the radiance for a set of parameters that is consistent with 
 observations. The results are then statistically compared to the 
observational data. 

The main aims of this paper are \\
I. to identify diagnostics which allow such a statistical comparison 
to be made. The two diagnostics found and studied are the lognormal 
probability density function (PDF)   
of EUV radiances and the wavelet or Fourier power spectrum 
of radiance time series; \\
II. to test to what extent the use of these diagnostics can constrain the parameters of the 
model, in particular the power-law exponent  $\alpha$ of the distribution of 
nanoflare amplitudes; \\ 
III. to compare with SUMER transition region and coronal data in order to check whether this simple 
model can reproduce the observations and whether the deduced parameters are 
realistic. \\ 
We begin with a description of the data and their analysis (Sect.~2). 
Next we give an outline of the model used for the flare simulation 
and establish a heuristic     %
reasoning for a lognormal distribution of 
the radiances  under the assumption that they are entirely due to 
transient events (Sect.~\ref{sect_mod}). 
The results of parameter studies with our model are presented in Sect.~4.  
In Section~5 we compare the simulations with SUMER 
measurements, 
and we outline our conclusions in Sect.~6.

\section{The SUMER data: reduction and analysis}

\subsection{Data and reduction}
SUMER is a stigmatic normal incidence telescope and spectrometer, operating 
in the wavelength range from 46.5 to 161.0~nm, depending on the spectral 
order and the choice of detector.
For a general description of the SUMER instrument and its data 
we refer to  \cite{wilhelm}.
The SUMER slit with angular dimensions of 1\arcsec$\times$300\arcsec\ 
used for the data analyzed here 
is imaged by the spectrograph on to the detectors with a 
resolution of about 1\arcsec\ per pixel 
in the spatial direction and 4.4~pm   %
per spectral pixel in first order and 2.2~pm %
per spectral pixel in second order. 

The \oiv line at 79.0~nm 
and the \ne8 line at 77.0~nm, 
which are used here, are measured in first order. 
We also analyzed and modelled the \siv\ line at 139.3~nm, but the results 
were very similar to those pertaining to the \oiv line and are not described 
in detail in the following. 
On 8 February 1998,  SUMER  performed 
a long-duration observation of quiet areas near disk centre in the 
 \oiv  and \ne8 lines. 
The measurements, which have been described by \cite{wilhelmkalk}, 
were taken over 7~h and 20~min 
with a cadence of 33.5~s 
and a particularly accurate compensation of the solar rotation. 
 The \siv\  measurements, recorded 
19 July 1998, belong to a quiet Sun explosive 
events study. Recordings were made over 3~h and 35~min with a cadence of 
15~s.   %
(See Table~\ref{tab1} for a summary of the measurements.) 

The data were corrected for the flatfield, the geometric distortion,  
and for detector electronics effects such as dead-time and local-gain 
depression. 

After the instrumental corrections and the radiometric calibration, 
the solar radiances were determined by integration over the line profiles,
which were derived by least-squares fits of single %
Gaussian functions and a linear background. 
The spectral background (continuum) was subtracted prior to integration. 

The data used by \cite{pauluhn00} were mainly image scans and thus 
gave the distribution of the radiances in the scanned area 
during the time span of the image recording. 
They may be considered to give a snapshot of the distribution of radiances. 
Here, we investigate time series, i.e., we 
focus on the distribution of radiances in a certain area caused by 
temporal variations. 

\subsection{Data analysis: lognormal distribution}

Before doing any modelling we 
investigate whether  time series of quiet Sun radiances are approximated by a lognormal 
distribution.  
After finding that this is the case,  we  
determine the parameters of the distribution,  which is characterised  
by the following probability density function   $\rho$ for an independent 
parameter $x$ (e.g., the radiance) 
\begin{equation} \label{eqlognormal}
 \rho(x)=\frac{N_{0}}{\si x \sqrt{2\pi}} \exp(-\frac{(\log(x)-\mu)^2}{2 \si^2}), 
\end{equation}
with $\mu = <\log(x)>$, $\si = \sqrt{Var(\log(x))}$ and $N_{0}$ a 
normalization factor. $Var$ means the variance. 
$\si$ is called the shape parameter because it determines the shape of 
the distribution: small $\si$ makes the distribution more Gaussian-like, large $\si$ 
makes the distribution more skewed. 
$\mu$ is called the scale parameter and stretches the distribution function. 

\begin{figure}[h]
\unitlength1cm
\includegraphics[width=0.48\textwidth,clip=]{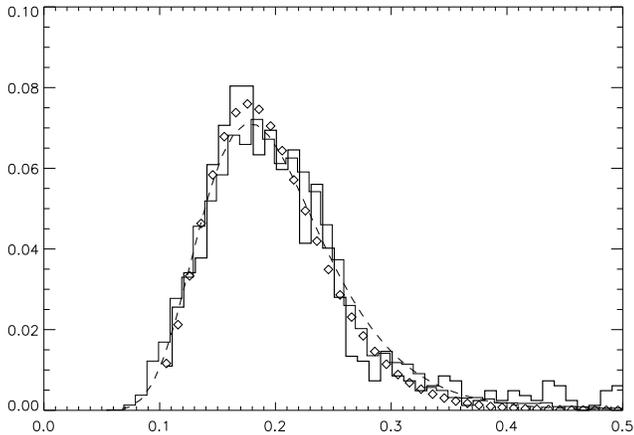} 
\caption[1]
{Example of a histogram from a SUMER transition region time series   
(quiet area, \siv\ line at 139.3~nm). The radiance data and the corresponding 
lognormal fit are 
given by the thick-line histogram and diamonds, and a simulation and its fit are indicated  
by the thin-line histogram and a dashed curve. }
\label{fig_Si4_tr_cells}
\end{figure}

 An example histogram of a SUMER time series of the transition region 
(\siv\ line at 139.3~nm) is shown in Fig.~\ref{fig_Si4_tr_cells}, along with a lognormal fit.   
The bin size is  5\% of the maximum radiance value. 
 For practically all spatial pixels 
the distributions of the O\ts{$\scriptstyle {\rm IV}$, 
the \siv, and the \ne8 radiances are well described by  
lognormal functions.  %
This does not automatically follow from the result of \cite{pauluhn00}, 
since they considered the distribution of radiance from different spatial 
locations, which represented a mix of network and cell interior locations. 
Also,  the lognormal distributions from network and cell deduced from 
the time series differ in their   
$\mu$ values, but not significantly in their $\sigma$ values  
(see Sect.~\ref{sumanalysis}). 

At coronal temperatures the shapes of the probability density functions are 
narrower than for the transition region and more symmetric 
(cf., Pauluhn et al., 2000), %
and the time series shows peaks that are less well marked, which will be shown in 
Sect.~\ref{sumanalysis}. 
Thus the difference between the distributions of radiances reflecting 
different temperatures is equally shown by data sampling a part of the 
solar surface or a single point over a time interval. 
The  $\mu$ and $\sigma$ values found from fits to the observations are therefore 
useful constraints to any model of quiet-Sun radiances. 
\begin{table*}
\caption{The SUMER data used in this paper.}
\label{tab1}
\begin{center} 
\begin{tabular}{ccccc} \hline
line & region & date  & duration/cadence & remarks \\ 
\hline 
\hline
\oiv  79.0 nm & TR, $T\approx 10^5$~K & 8 February 1998 
 & 7 h 20 min, 33.5~s & quiet Sun study, high-telemetry \\
\ne8  77.0 nm & upper TR/lower corona $T\approx 7\times 10^5$~K 
& 8 February 1998 & 7 h 20 min, 33.5~s  & quiet Sun study, high-telemetry\\
\siv  139.3 nm & TR, $T\approx 10^5$~K & 19 July 1998 & 3 h 35 min, 15~s 
& quiet Sun explosive events study \\
\hline
\end{tabular}
\end{center} 
\end{table*}

\section{A simple model} \label{sect_mod}

\subsection{Description}
Using a  simple model of %
transient brightenings, we produce synthetic  
time series of EUV radiances. We presume that flaring is an 
intrinsically stochastic process, and our radiance variable is a time-dependent
 random variable.  %
One simulation thus delivers a possible realization of this process. 

Our model consists of a time series of random kicks 
(acting as ``flares''), applied to an initial radiance. 
Each kick is followed by the exponential decay of 
the radiance. The final radiance is given by the sum of the radiances of all 
the overlapping transient brightenings. 
The numerical value of the initial radiance is 
not important, since it is damped just like every brightening and  
the model reaches a statistical ``steady state'' after 
a certain relaxation time. 
We have 5   
free parameters: the maximum and minimum allowed flare amplitude, $y_{max}$ and 
 $y_{min}$,  the power-law exponent $\al$, 
the e-folding or damping time of the flare $\tau_d$, 
and the frequency of the excitation, i.e., 
the flaring rate or flaring probability $p_f$. 
This set is complete if we assume that the shape of the nanoflare is given 
by a single kick with a 
sharp rise and a successive exponential decrease. 
These parameters are sufficient to quantitatively reproduce the radiance PDFs and 
power spectra.
In order to smooth the steep increase a rise time $\tau_r$ 
can be introduced (which, if included, is chosen to be a fixed fraction of 
the damping time and hence does not add an additional free parameter). 
The introduction of a rise time improves the qualitative agreement between 
synthetic and observed time series. 

For the energy content of a flaring event 
of amplitude $y_{0}$  and damping time $\tau_d$   the following proportionality holds 
 \begin{equation} \label{eqenergy0}
 E \sim \int_{t_0}^{\infty} y_{0} e^{-\frac{(t-t_0)}{\tau_d}} dt = 
y_{0} \tau_d .
\end{equation}
 
Thus,  
the energy content of a flare with amplitude $y_{0}$, damping time  
$\tau_d$ and rise time $\tau_r$ 
can be estimated as 

\begin{equation} \label{eqenergy}
 E = q y_{0} \pi A (\tau_d + \tau_r), 
\end{equation}
 where $A$ is the solar surface  
 area, if we assume equal radiation in all directions (Lambertian 
surface), and $q$ is a fraction of the total energy radiated in the 
observed spectral line. In general $q$ is a small number, determining 
whose exact value is beyond the scope of the present paper. 

\subsection{The simulation}

The model produces realizations of possible radiance time series, i.e., 
we get the radiance as a time-dependent 
stochastic variable (stochastic process), and it describes 
an example of a process which is close to being Markovian.  
In a Markov process, the stochastic variable 
at one timestep $t_{n+1}$ is only dependent on the directly 
preceding timestep $t_{n}$. It has ``nearly no memory'' of the history,  
and the probability to find a variable at position $x_{n+1}$ at time 
$t_{n+1}$ is calculated by an initial probability and the two-time 
transition probability (which is the conditional probability that the 
variable is in state  
$x_{n+1}$ at $t_{n+1}$ under the condition that it has been in state 
$x_{n}$ at $t_{n}$).

The simulation involves the following steps:

\begin{enumerate}

\item Generate a distribution of flare strengths, i.e. (positive)  
values of flare amplitudes $f_{n}$. 

\item Start from an initial radiance value $r_{0} > 0$  
(the ``first kick'', taken from the flare distribution, $r_{0}=f_{0}$). 

\item At random time $t_{i}$ another radiance $r_{i}$ is generated 
by adding a flare kick $f_{i}$,  

\begin{equation}
r_{i} = r_{i-1}+ f_{i}. \end{equation}
The random amplitude $f_{i}$ is required to be consistent with the chosen 
distribution function. 

\item At successive time steps $t_{j}$, $j > i$, the radiance values are  
 \begin{equation} \label{expondamp}
r_{j} = r_{i} \cdot \exp{(-\frac{t_{j}-t_{i}}{\tau_{d}})},\end{equation}
with $\tau_{d}$ the damping time,  
corresponding to a difference equation of 
 \begin{equation} \label{expondamp_diff}
r_{j}-r_{j-1}= r_{j-1} ( \exp{(-\frac{t_{j}-t_{j-1}}{\tau_{d}})} - 1),
\end{equation}

 multiplicatively generated from the preceding values. 
Here we assume for simplicity that all brightening events have the same 
damping time.

\end{enumerate}
The probability of a transient brightening occurring (per time step), 
$p_f$ with  $0 < p_f < 1$, is simulated by 
drawing equally distributed random numbers between 0 and 1, and a flare event 
is started at $t_i$ if the random number falls within a certain fraction 
of the interval (0,1). For example, a flaring probability of 30~\% or 0.3 is realized 
by applying the kick if the random number is smaller than 0.3. 
The flare process is thus a  Poisson process, which is equivalent to the fact 
that the waiting times, i.e., the time intervals between two flares 
$\Delta T_i$, 
are exponentially distributed with parameter $p_f$, and the mean waiting time 
$<\Delta T>$ is given by $1/p_f$. Additionally to the excitation, we include 
a damping process  with damping time $\tau_d$, which is inverse to the 
strength of the damping process. %
In Fig.~\ref{fig_singleflares} an example sequence of flares with very large 
mean waiting times or low frequency $p_f$ is shown.
For higher frequencies, the radiances due to individual ``kicks'' 
or flare-like events overlap much more.

\begin{figure} %
\unitlength1cm
\begin{center}
\includegraphics[width=0.5\textwidth,clip=]{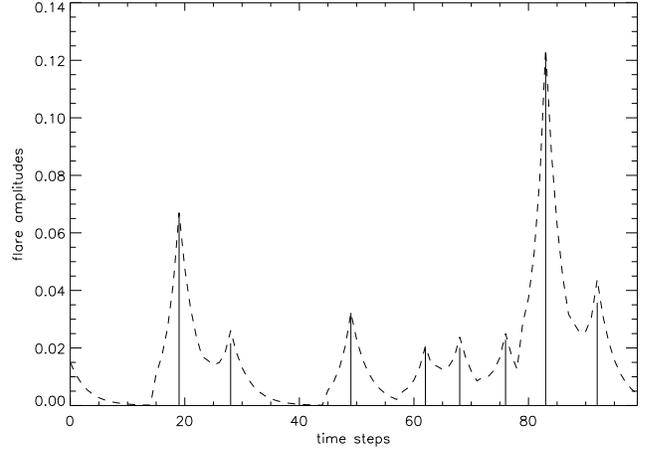}
\end{center}
\caption[1]
{An example sequence of simulated flares. Here, the flaring frequency has been 
chosen to be very small, $p_f=0.08$,  %
such that the average waiting time between 
 two flares is 12.5 time steps.} 
\label{fig_singleflares}
\end{figure}
A heuristic explanation for the distribution of the radiances is given in the 
following. 
We assume that we can express the kick as   
   $f_{i}=c \cdot r_{i-1}$, with $c$ %
a positive value, i.e., the new radiance is 

\begin{equation}
r_{i} = r_{i-1}+ c \cdot r_{i-1} = \hat{c}\cdot r_{i-1},\end{equation}
a multiple of the predecessor radiance.
We stress, however, that we prescribe the distribution of the additive 
components $f_i$ (given, e.g., by observation as a power law) and not that 
of the multiplicative component $c$.

Thus, the radiance value at the $n$th step is 
\begin{equation} \label{eq5}
 r_{n} = r_{0}\cdot c_{1} \cdot c_{2}\cdot \cdots \cdot c_{n},\end{equation}
with the  
$c_{i}$ being random  %
factors if a new kick happens at this time. Otherwise $c_i$ represents the 
damping as described by the exponential in Eq.~(\ref{expondamp}).  

This assumption of a multiplicative process is plausible in the light of the 
radiance being ultimately generated by emerging and decaying magnetic 
field.  %
This is for example implied by the excellent correlation between high-resolution 
magnetograms and images in EUV and X-ray wavelengths, although the underlying 
processes are still far from being understood. 

Taking the logarithm of Eq.~(\ref{eq5}) yields
\begin{equation} \label{logsum}
\log(r_{n}) = \log(r_{0})+\log(c_{1})+\log(c_{2})+ \cdots + \log(c_{n}).
 \end{equation}

The Central Limit Theorem %
then states that 
as $n$ goes to infinity, the distribution of the sum converges to a normal 
distribution. 
For the  Central Limit Theorem to hold, the $c_i$ have to be 
independent random variables,   %
which is not precisely 
fulfilled here, because the flaring process is coupled via the damping 
time scale 
and thus has a (short) memory. We can, however, assume that the damping time 
is short relative to the other time scales of the system, e.g., the duration of the 
measurements, which would link the flare process  
to a coloured noise process instead of  
a white noise process 
\citep[see, e.g.,][]{honerkamp}. %

Thus, the distribution of the sum in Eq.~(\ref{logsum}) 
is approximately normal, so that the distribution of the 
$r_{n}$ is approximately lognormal. 
On the time scales under study (several hours), we  can assume 
stationarity of our 
distribution, which means that (after an initial relaxation time) 
a steady state is reached. This does  not hold for the solar radiance 
when the structure of the region changes significantly, 
e.g., due to emergence of new flux and thus a change in activity. 
Our study is thus limited to truly quiet regions. 

The two characterizing parameters of the stationary solution
Eq.~(\ref{eqlognormal}), $\mu$ and $\sigma$, 
have been determined empirically by parameter scans for Gaussian  
and for power-law stochastic flare input.

\section{Parameter studies}

In this section we describe the effects of varying 
the free parameters of the simulation on 
the parameters of the radiance distribution. The simulated  
 time series consisted of $n=5\times10^5$ time steps each. 

First we need 
to demonstrate that the distribution 
function of the radiance has a lognormal shape. 
We ran the model 
for 
flare amplitudes which are distributed according to a power law, with 
the exponent $\al=2.1$ and the amplitude ranging  
between 0.02 and 3.0 W\,m$^{-2}$sr$^{-1}$. 
Figure~\ref{fig_example2} shows examples of simulated distributions with 
damping times $\tau_d = 10, 50, 100$ and 200 time steps, and a flaring 
frequency of $p_f =0.2$ (in inverse time steps). 
Clearly, the resulting distributions are highly skewed towards small 
radiances for a short damping time. In the limit of infinitely short 
damping time the most common value of the radiance is zero. 
As the damping time increases, so do the average value 
and the width of the radiance distribution, which also becomes 
increasingly symmetric. 
The same behaviour is found as the flaring frequency increases  (not plotted).
Note that for increasing $\tau_d$, as the overlap between individual 
brightenings increases, a background of nearly constant brightness is 
built up. This becomes visible in Fig.~\ref{fig_example2} through the 
increasing amount of points at small radiances at which the PDF is zero.

In order to check if a lognormal results only for a power law we also 
considered, as a test case, a Gaussian 
distribution of flares, which results in a far more symmetric 
distribution of radiances than when the distribution of flare amplitudes 
is a power law. 
Nevertheless, it is also well approximated by a lognormal distribution. 
In the following we consider only the power-law distribution of flare 
amplitudes, since it is supported by observation.  
\begin{figure} %
\unitlength1cm
\begin{center}
\begin{picture}(10,6.)
\put(1,4.9){$\tau_d=$10}  
\put(1.6,4.){$\tau_d=$50} 
\put(2.5,2.9){$\tau_d=$100}
\put(4.5,2.2){$\tau_d=$200}
\includegraphics[width=.45\textwidth,clip=]{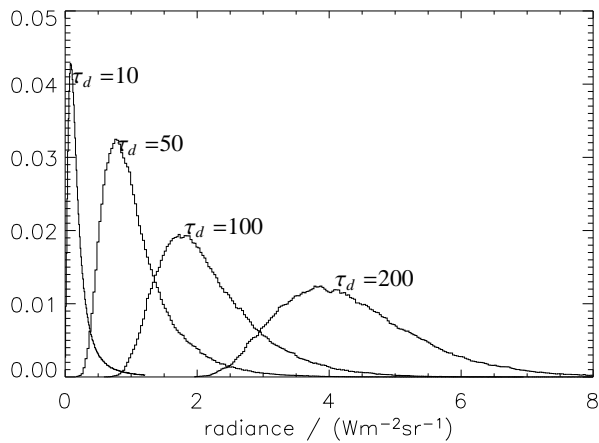}
\end{picture}
\end{center}
\caption[1]
{Histograms of the simulations with $\tau_d=$10, 50, 100, and 200 time steps 
for  power-law flare distributions with fixed flaring probability of 0.2.} 
\label{fig_example2}
\end{figure}

\subsection{Variation of the damping time and flaring frequency}

The shape of the equilibrium distribution is generally 
determined by the ratio of damping to excitation.  
Only when the two forces balance each 
other, can a stationary state be reached. 
To confirm this for our simulations, we vary the quotient of damping 
strength and excitation frequency $\frac{1}{\tau_d p_f}$ in two 
different ways: first, we vary $\tau_d$ and allow $p_f$ to remain constant. 
With increasing damping time, the shape parameter $\sigma$ decreases, 
which means that the skewness 
of the lognormal decreases. The distribution becomes more symmetric 
when the damping is weaker. 
Then we do the opposite and vary $p_f$ while leaving  $\tau_d$ unchanged. 
The results (for the parameters  
$\al=2.5, y_{min}=0.016, y_{max}=0.8$)  are shown in Fig.~\ref{fig_sigwithratio}.
Both curves agree with each other within the uncertainties of the fits.  
From Fig.~\ref{fig_sigwithratio} it  follows 
\begin{equation}\label{sig}
 \sqrt{2}\sigma   = \frac{K_{\alpha}}{\sqrt{\tau_d p_f}} + const.  
\end{equation}
The proportionality factor $K_{\alpha}$ varies between values of 1  and 2.
As either $\tau_d$ or $p_f$ increase the radiance enhancements due to consecutive 
kicks  increasingly overlap, leading to an increasingly symmetric distribution.
For very low $\tau_d$ and $p_f$ low radiances dominate the distribution, while 
the few high peaks make it very asymmetric (see Fig.~\ref{fig_example2}). Also empirically, 
we find the following 
dependence for $\mu$:
\begin{equation}\label{mu}
\mu=\log(\tau_d m_f p_f) + p_f \exp(p_f) + const., 
\end{equation}
where $m_f$ is the mean value of the flare amplitudes.
\begin{figure}[h]
\unitlength1cm
\begin{picture}(.4,0.1)
\put(0.1,3.5){$\sqrt{2}\sigma$}  \put(4.6,-0.2){$(\tau_{d}p_f)^{-1}$}
\end{picture}
\centerline{\hbox{
\includegraphics[width=0.45\textwidth,clip=]{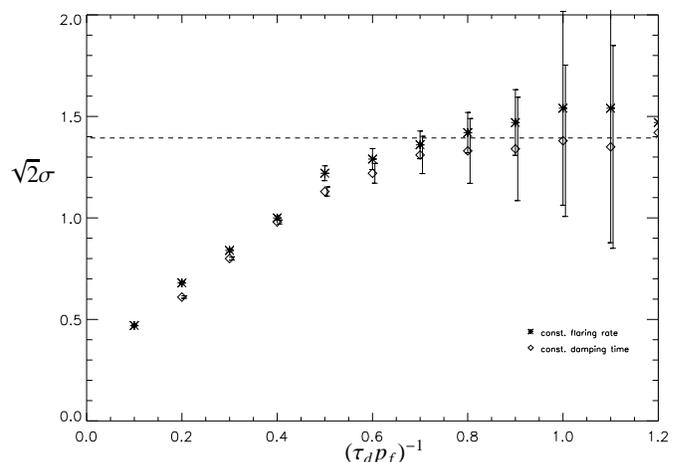}
}}
\caption[1]
{Variation of the lognormal shape parameter with the ratio of damping to flaring rate.
The stars give the results for constant flaring rate $p_f$ and varying damping time 
 $\tau_d$,      %
and the diamonds give the results for constant $\tau_d$  and varying $p_f$.
The horizontal dashed line represents the observed value as deduced from SUMER data.}
\label{fig_sigwithratio}
\end{figure}

Where the damping or excitation %
exceed certain limiting values, no 
stable solution exists that can be described by a lognormal distribution. 
If the excitation is too strong (for this set of parameters at $\frac{1}{\tau_d p_f}\le 0.1$),  
the realizations grow unboundedly, and %
at the other end of the range,
if the product $\frac{1}{\tau_d p_f}$ exceeds a certain value (here 1.2), 
the shape of the radiance distribution 
changes from a two-sided function to a function with a maximum value at or very near 
zero and a rightwards extending tail.

\subsection{Dependence on the minimum and maximum flare amplitude}

For constant flaring frequency and damping  %
time the energy input 
to the system is determined by the flare distribution, i.e., the minimum
and maximum flare amplitude and the exponent of the power law. 
For fixed upper and lower boundaries, the energy input decreases with 
increasing exponent. Thus, in order to reproduce a certain (measured) radiance 
time series and its mean value, the  boundary $y_{min}$ and/or $y_{max}$
has to be varied 
simultaneously with the exponent such as to keep the energy input 
constant. 
The mean of the flare radiance input is given by 
\begin{eqnarray}
 m_{f} &=&E_{0} \int_{y_{min}}^{y_{max}} E E^{-\al} {\rm d}E = 
 E_{0} \int_{y_{min}}^{y_{max}} E^{1-\al} {\rm d}E \\ 
  & \rm{with} & \quad E_{0}= 
   \left( \int_{y_{min}}^{y_{max}} E^{-\al} {\rm d}E \right)^{-1}   \nonumber  \\
        &=& \frac{(1-\al)}{(2-\al)}\frac{(y_{max}^{2-\al} - y_{min}^{2-\al})}{(y_{max}^{1-\al} - y_{min}^{1-\al})} . \nonumber
\label{eq_flareenergy}
\end{eqnarray}
Figure~\ref{fig_varyminmax} shows the dependence of the lognormal parameters on 
the minimum and the maximum flare amplitude for $\al=1.8$.  
We find that the distribution of the radiances depends more strongly 
on the lower energy limit of the nanoflares than on the upper limit also for 
other $\al$ values, including $\al>2$.  
The mean of the resulting radiance values varies as the mean 
of a lognormal function   
\begin{equation} \label{meanlognormal}
\rm{mean}_{logn} = e^{\mu +\frac{\sigma^2}{2}}.  
\end{equation}
\begin{figure}[h]
\unitlength1cm
\centerline{\hbox{
\includegraphics[width=0.23\textwidth,clip=]{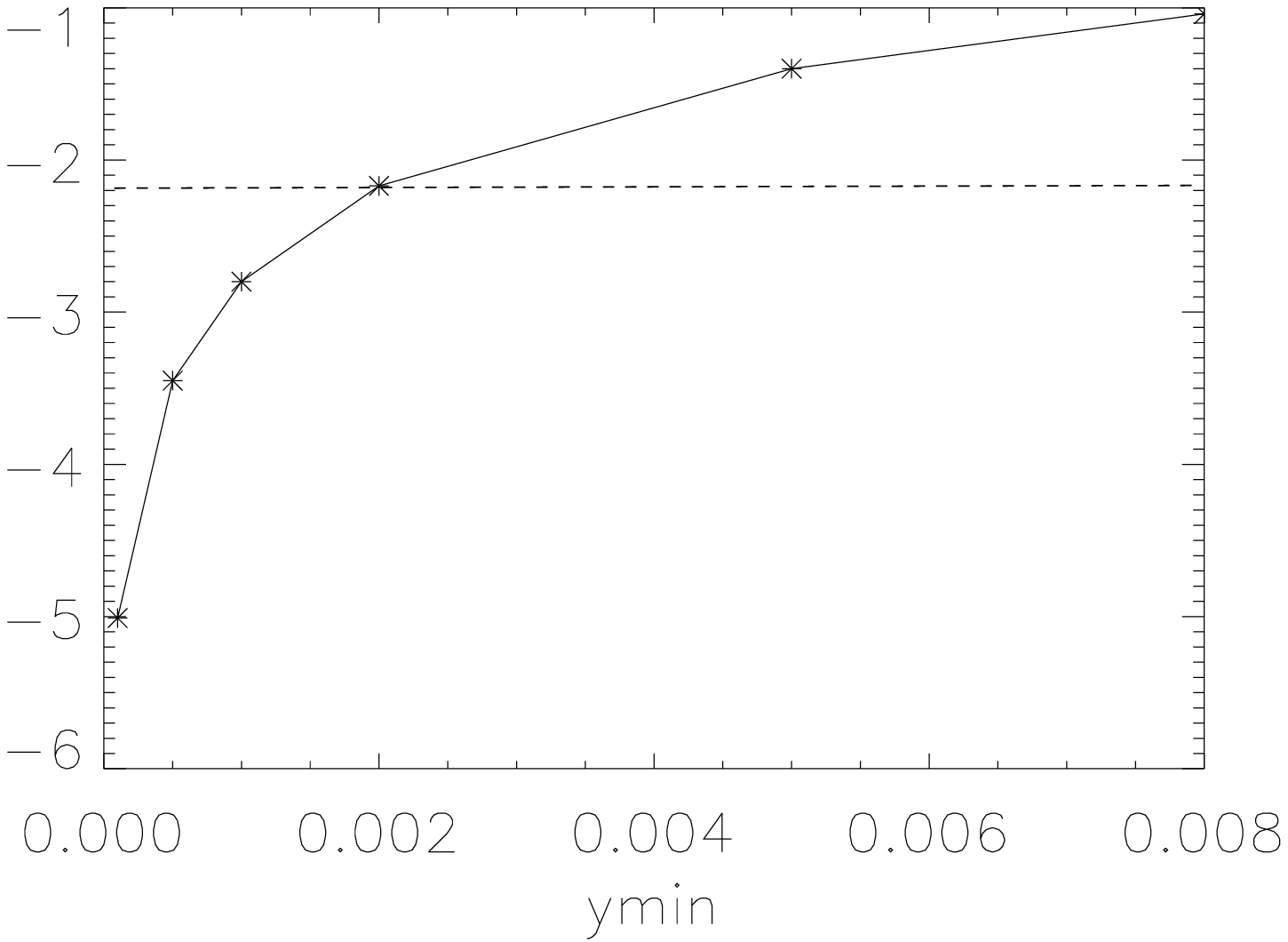}
\includegraphics[width=0.23\textwidth,clip=]{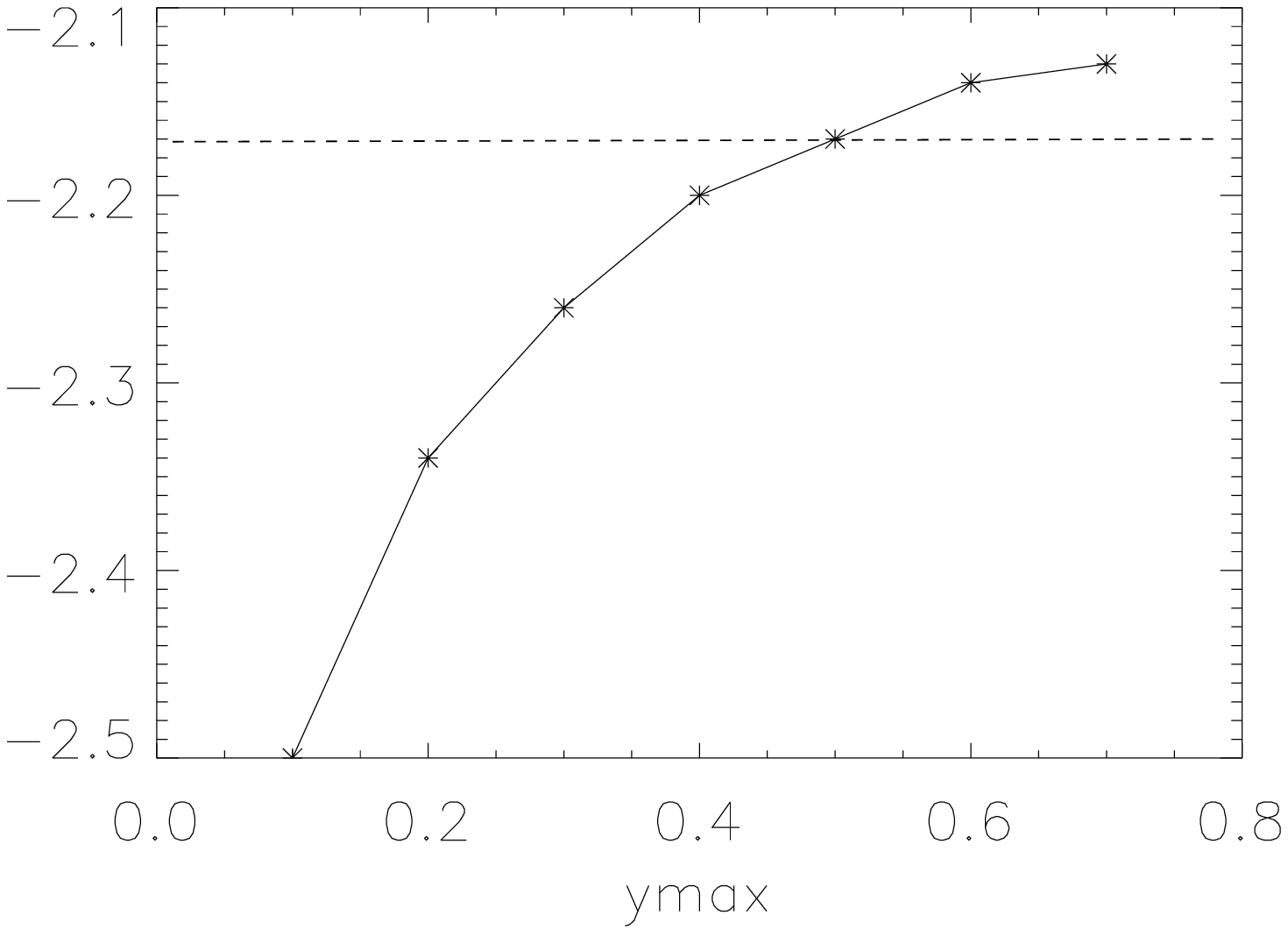}
}}
\centerline{\hbox{
\includegraphics[width=0.23\textwidth,clip=]{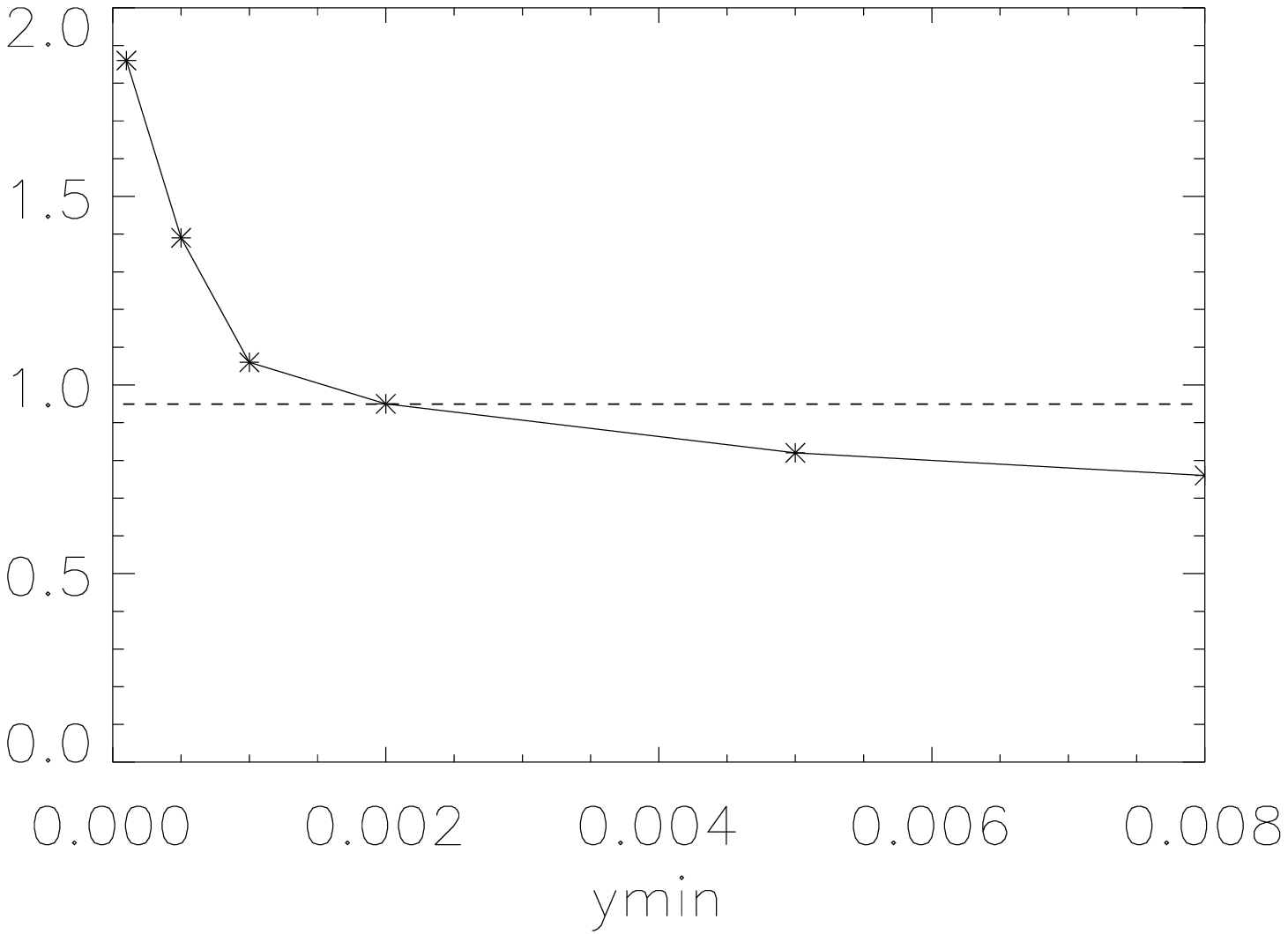}
\includegraphics[width=0.23\textwidth,clip=]{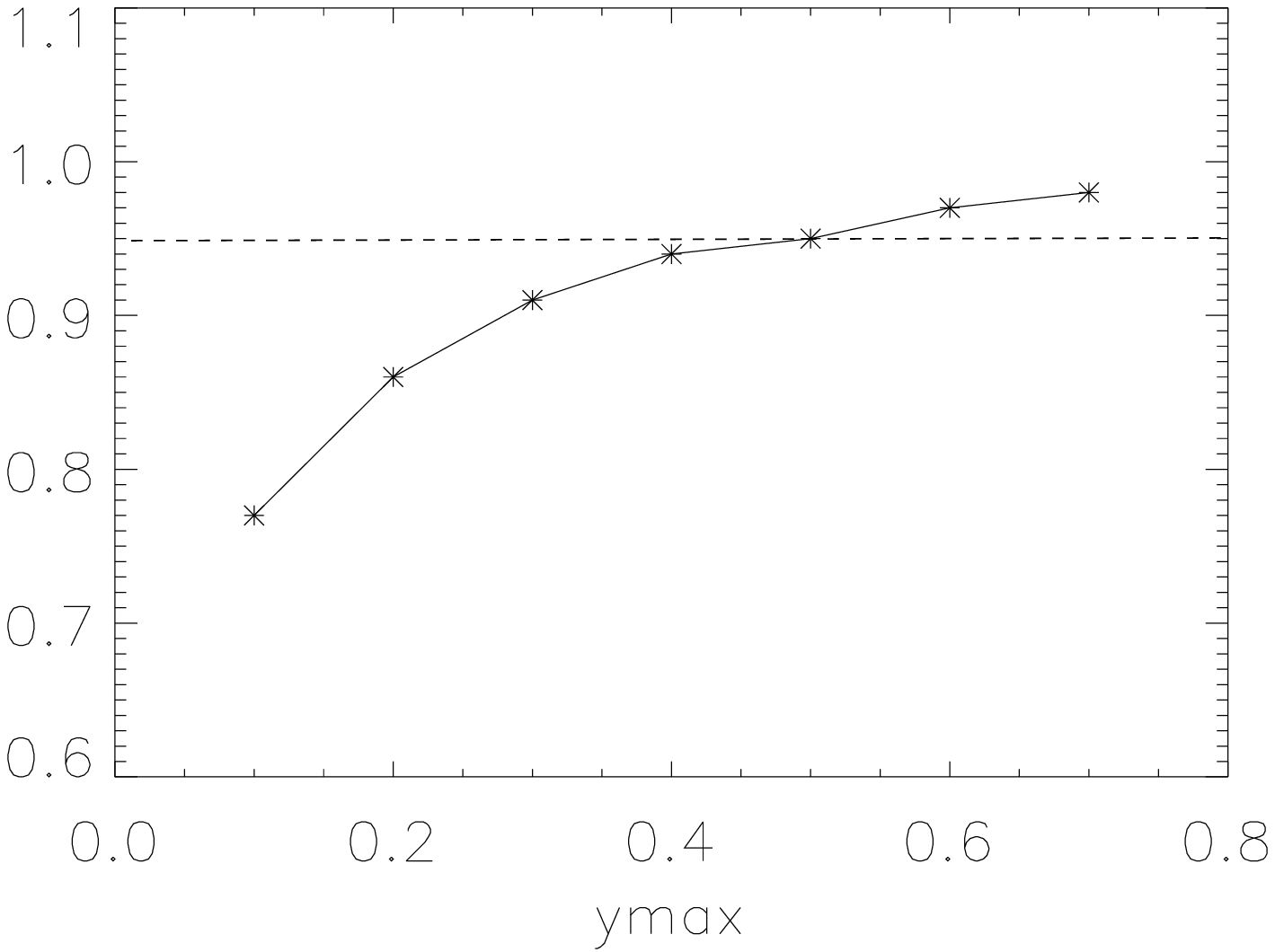}
}}
\begin{picture}(5,-1.)
\put(3.1,5.4){$\alpha=1.8$}  \put(7.4,5.4){$\alpha=1.8$} 
 \put(0.2,5.3){$\mu$}  \put(4.3,5.3){$\mu$}
 \put(3.5, 4.2){(a)}  \put(7.8,4.2){(b)}  

\put(3.1,2.2){$\alpha=1.8$}  \put(7.4,2.2){$\alpha=1.8$}
 \put(0.2, 2.1){$\sigma$}  \put(4.3,2.1){$\sigma$}  
 \put(3.5, 1.2){(c)}  \put(7.8,1.2){(d)}

\end{picture}
\caption[1]
{Variation of the lognormal scale and shape parameters with the lower 
(a,c) and upper (b,d) cutoffs of the flare amplitude distribution. 
The values employed for the other  parameters are $\al=1.8$, 
$\tau_d=8.01, p_f=0.60$,  
     ($y_{min}$ variable,   $y_{max}=0.5$);  and ($y_{max}$ variable, 
$y_{min}=0.002)$.  
The horizontal dashed lines represent observed values (averages over 
\oiv quiet network area). 
}
\label{fig_varyminmax}
\end{figure}
\begin{figure*}  %
\unitlength1cm
\begin{center}
\includegraphics[width=0.3\textwidth,clip=]{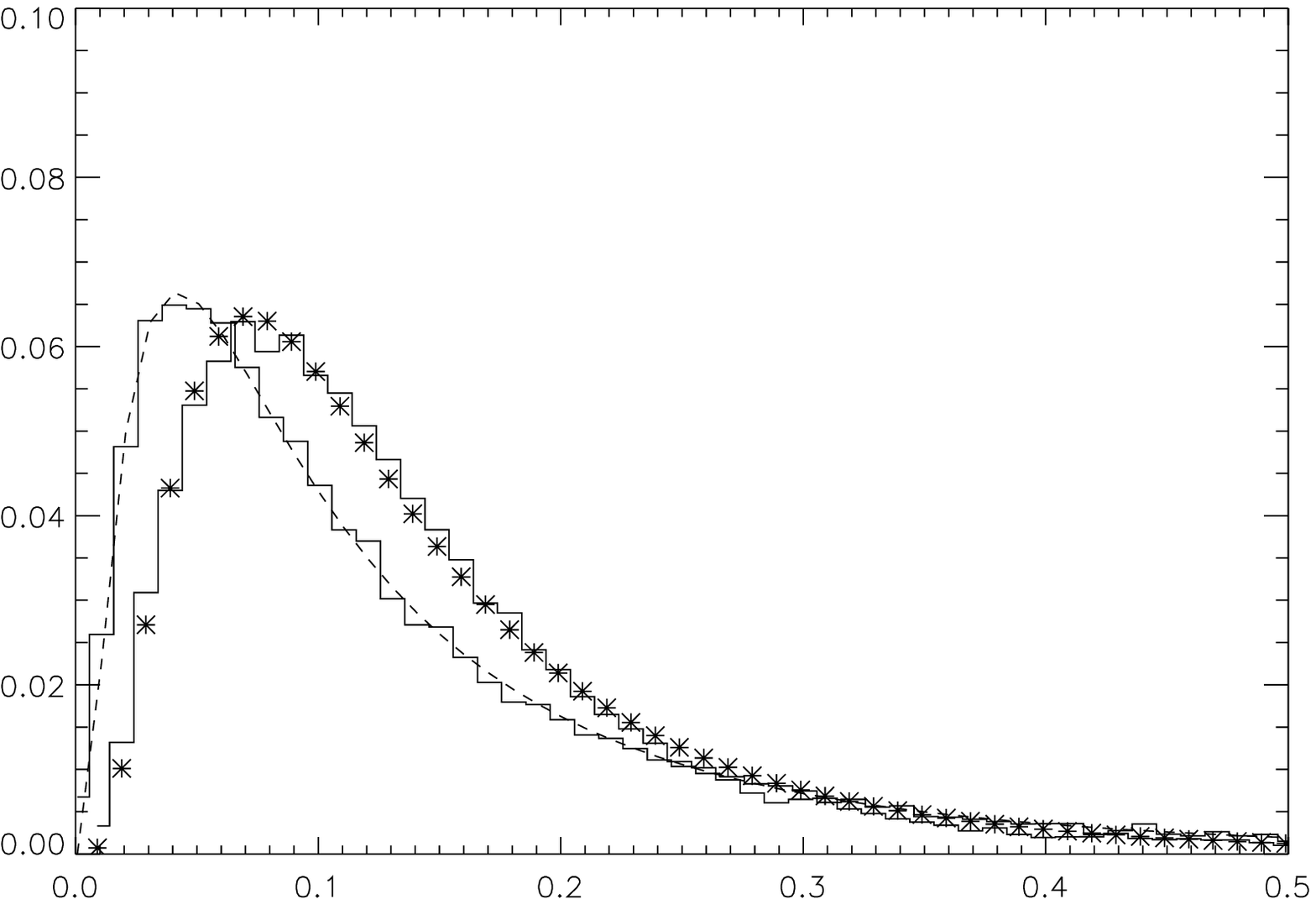}
\includegraphics[width=0.3\textwidth,clip=]{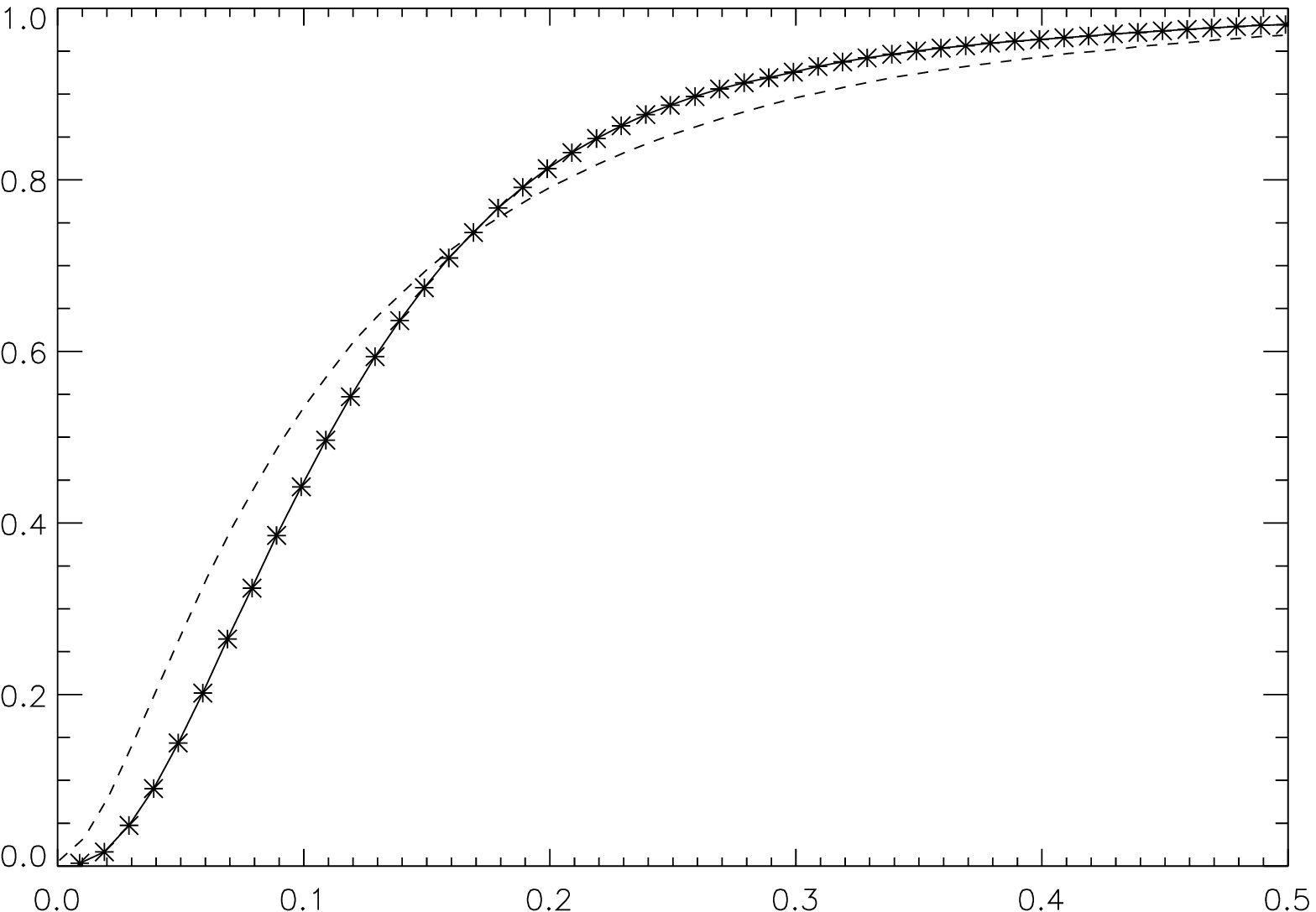}

\includegraphics[width=0.3\textwidth,clip=]{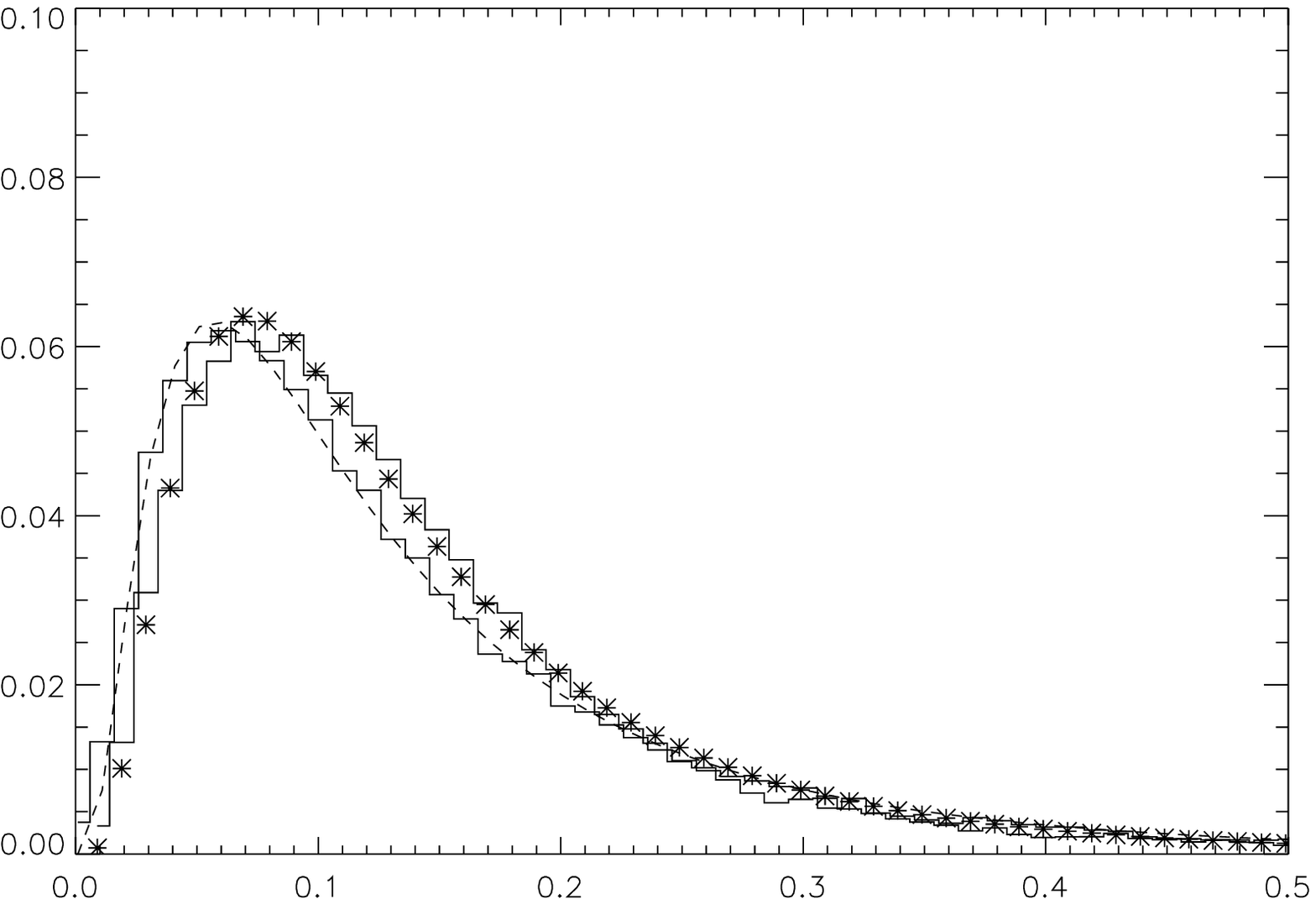}
\includegraphics[width=0.3\textwidth,clip=]{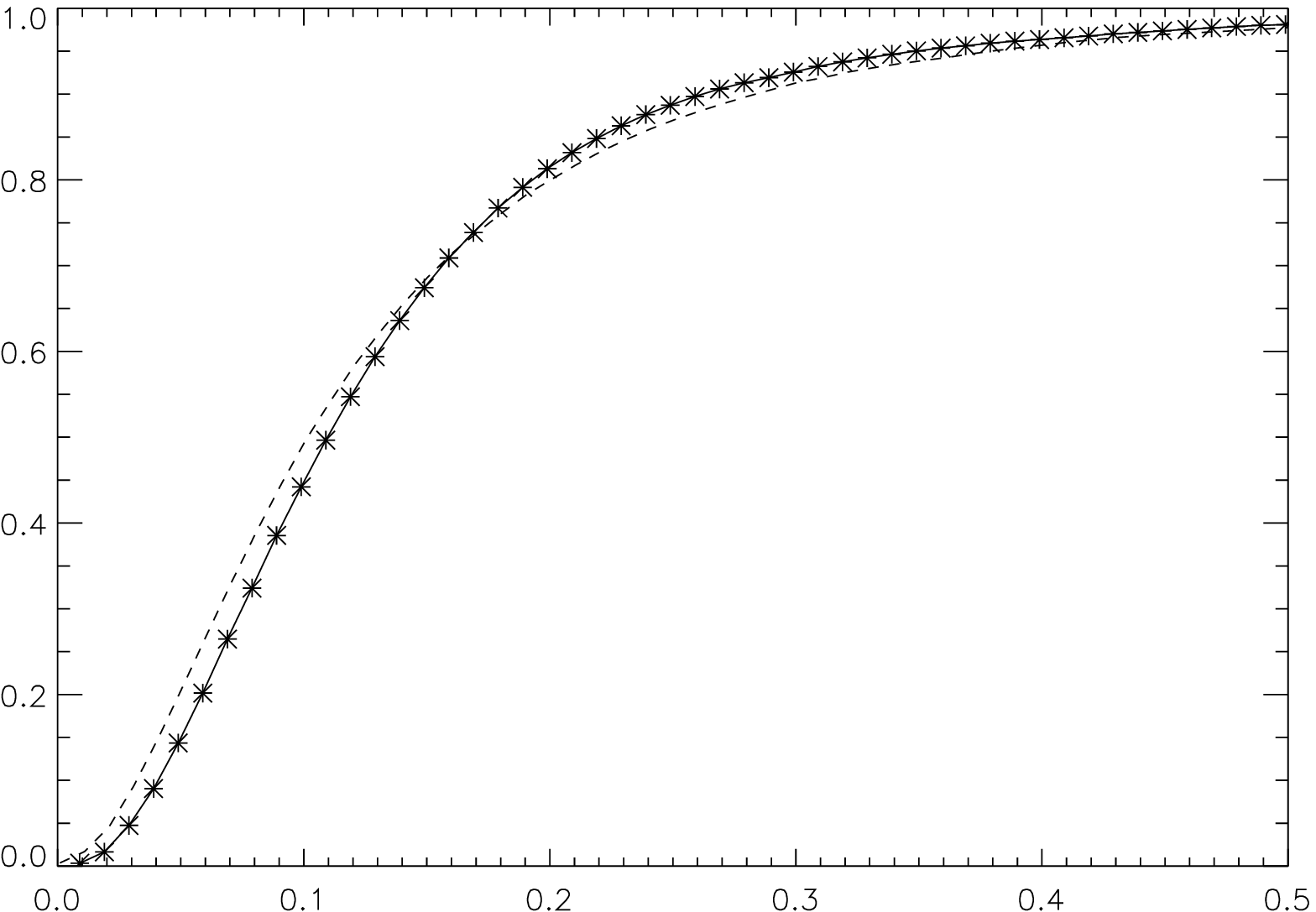}

\includegraphics[width=0.3\textwidth,clip=]{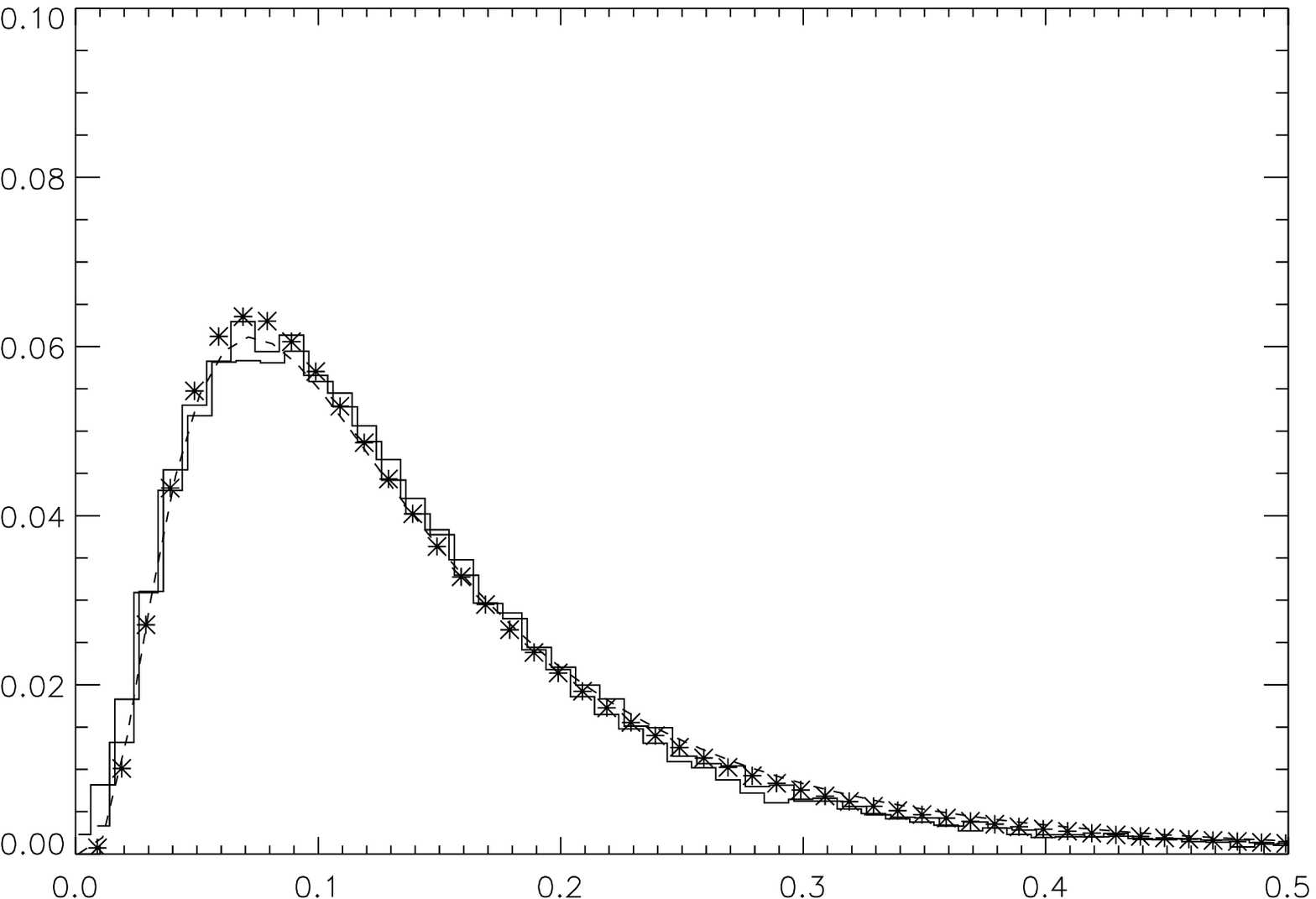}
\includegraphics[width=0.3\textwidth,clip=]{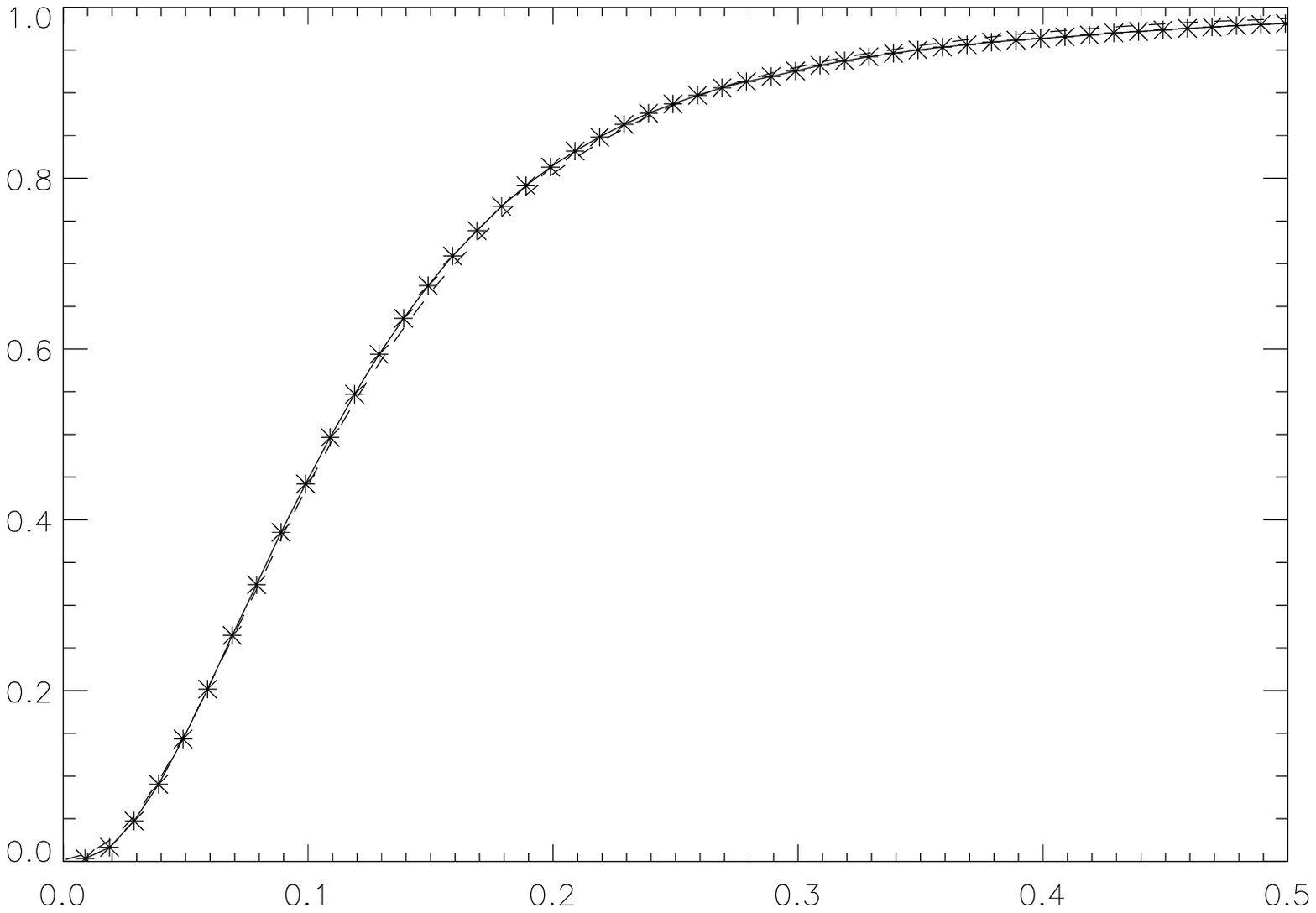}

\includegraphics[width=0.3\textwidth,clip=]{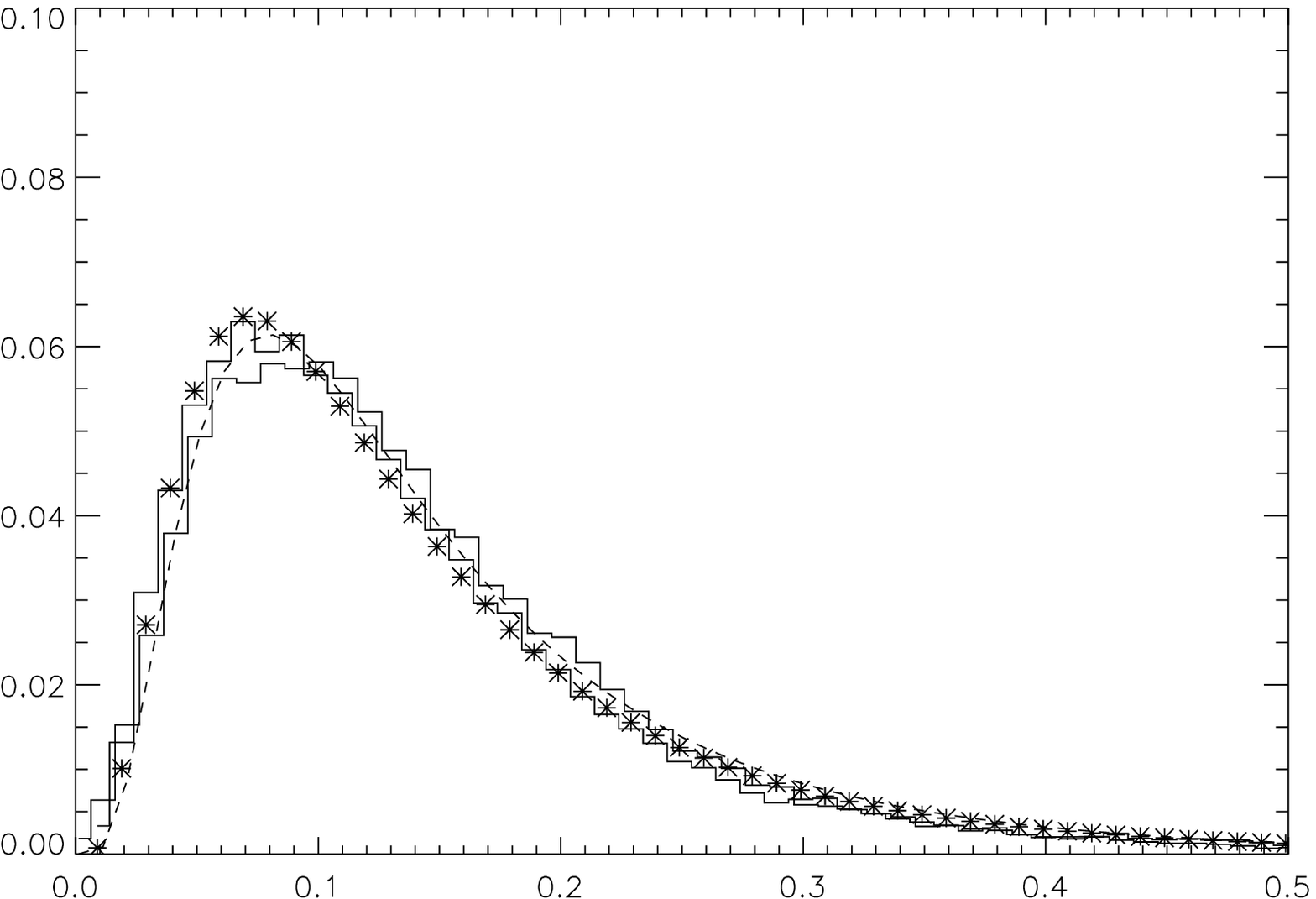}
\includegraphics[width=0.3\textwidth,clip=]{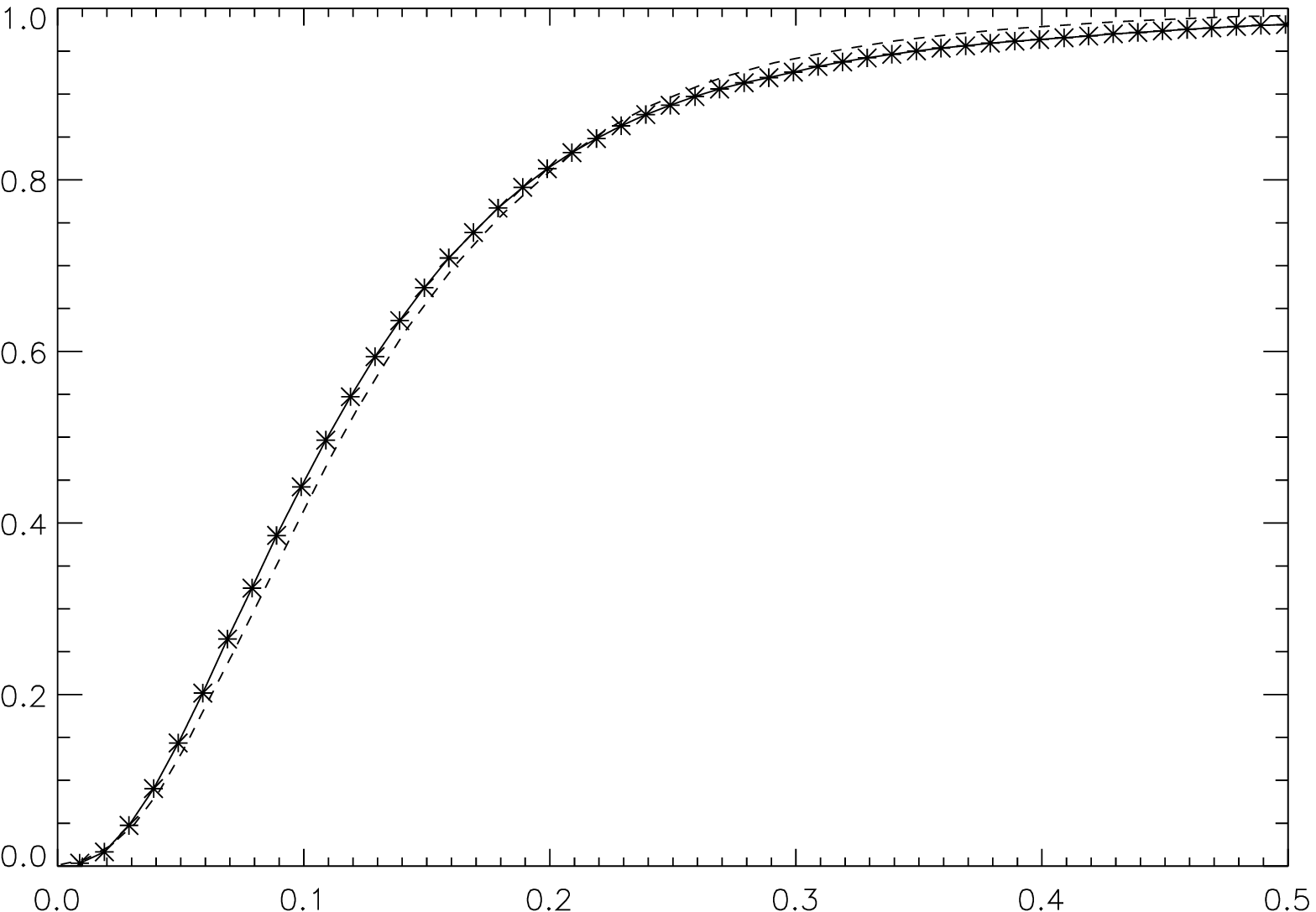}

\includegraphics[width=0.3\textwidth,clip=]{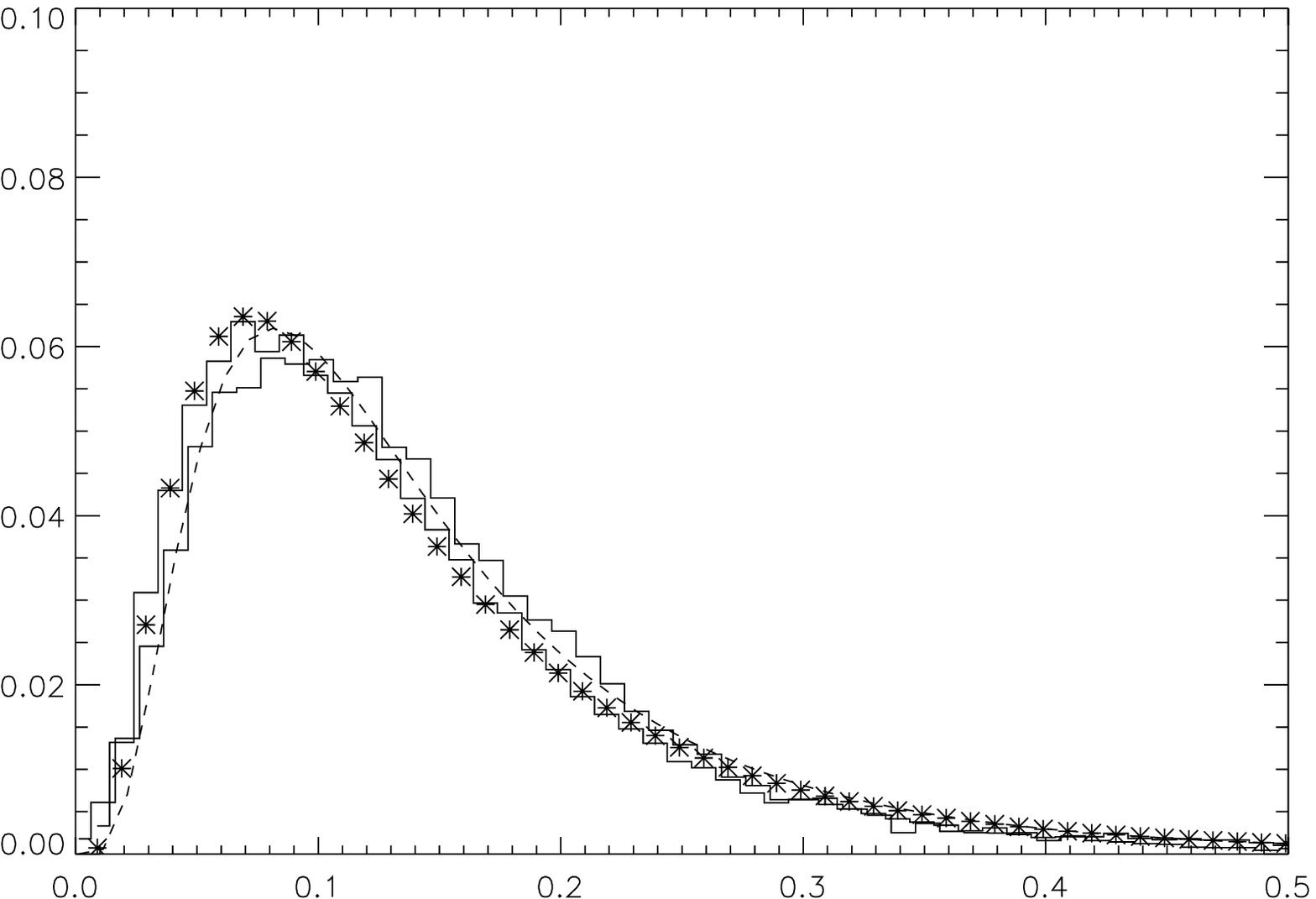}
\includegraphics[width=0.3\textwidth,clip=]{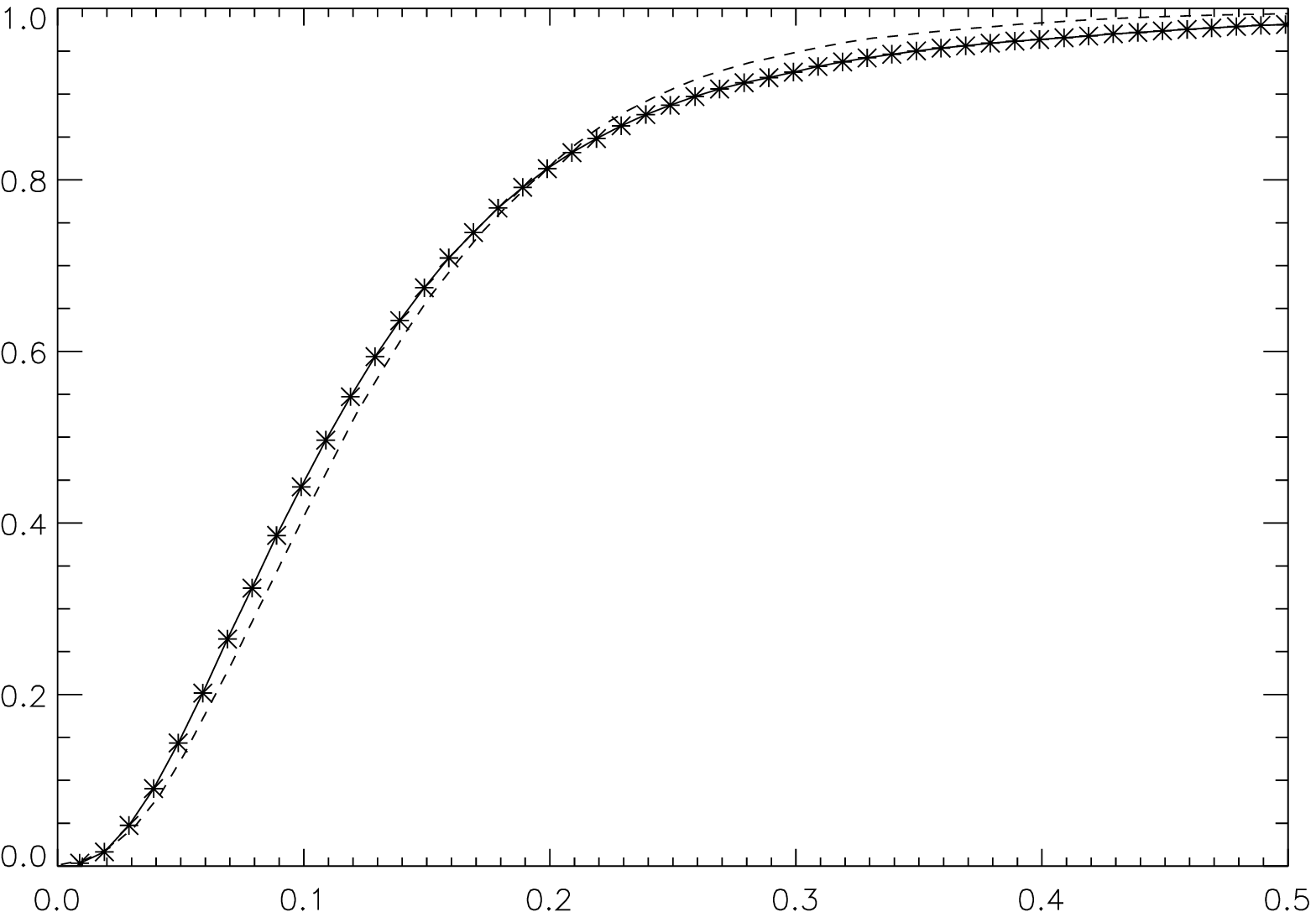}
\end{center}
\begin{picture}(18,0)
\put(15.,18){$\al=1.8$}
\put(15.,15){$\al=2.1$}
\put(15.,11){$\al=2.5$}
\put(15.,7.1){$\al=2.8$}
\put(15.,3.2){$\al=3.0$}
\put(3.4,17.){\begin{rotate}{90} Normalized frequency \end{rotate}} %
\put(14.8,18){\begin{rotate}{90} CDF  \end{rotate}}                 %
\put(5.1,0.5){radiance / W\,m$^{-2}$sr$^{-1}$}
\put(10.5,0.5){radiance / W\,m$^{-2}$sr$^{-1}$}
\end{picture}
\vspace{-0.5cm}
\caption[1]
{Variation of the radiance distribution (the radiance probability density 
function, shown on the left, and its cumulative distribution function, shown 
on the right)  with the exponent $\al$ of the 
flare distribution. %
The following parameters have been kept constant: 
$p_f=0.43$, $\tau_d+\tau_r=5.72$ time steps,   
and $y_{max}=0.8$. The simulated histogram is represented by the thin solid line. 
The dashed line indicates the lognormal-based fit to this distribution. 
Also shown are a SUMER \oiv histogram representative of the network 
 and the corresponding 
lognormal fit (stars).}
\label{fig_differentalph}
\end{figure*}
\subsection{Dependence on the power-law exponent} \label{sec_alphadep}

Figure~\ref{fig_differentalph} shows the probability densities and 
cumulative distribution functions (CDFs) 
of the simulated time series for different values of $\alpha$. 
Also plotted are the same quantities from the time series of the 
transition region \oiv 79.0~nm line recorded by SUMER in the quiet network. 
The histograms have been computed over the ensemble of all network time 
series in the observed quiet-Sun area  (60 pixels).   
They have been calculated with a binsize 
of 0.01~W\,m$^{-2}$\,sr$^{-1}$. %
The exponent $\alpha=2.5$ yielded the best match to the selected SUMER data. 
The quality of the fit was determined from a comparison of the density 
functions 
fitted to the histograms and the corresponding cumulative 
distribution functions.  
However, 
this diagnostic does not allow values of $\alpha \geq 2.5$ to be distinguished from 
each other unless the quality of the data is high and the statistics are good. 
In Fig.~\ref{fig_musigwithalpha} we show the lognormal parameters $\mu$ and $\sigma$ as a 
function of $\alpha$ for 
constant flare frequency $p_f=0.43$ ($1.3\times 10^{-2}$~Hz), damping time 
$\tau_d+\tau_r=5.72$ time steps 
(191.7~s) and constant upper amplitude value of $y_{max}=0.8$, 
the value of $\alpha$ has been varied 
while also varying the 
$y_{min}$ value ($y_{min}=$ 0.002, 0.006, 0.011, 0.016, 0.019, 0.021, respectively), 
such as to keep the energy input constant (as well as the number of flares).  
Clearly, the distribution of radiances becomes more symmetric 
  when $\alpha$ increases (when the amplitude range is selected in order to keep the 
mean value constant). %
It is also evident from Fig.~\ref{fig_differentalph} that the sensitivity of the radiance 
distribution 
to variations of $\alpha$  is highest for small $\alpha$ values. 
Figure~\ref{fig_musigwithalpha} confirms the result obtained from 
Fig.~\ref{fig_differentalph} that $\al=2.5$ agrees best with the SUMER data. 

\subsection{The power spectrum as an additional diagnostic} \label{sec_44}

Not surprisingly, just the probability density function alone does not 
lead to a unique solution, since the PDF contains no information on time scales. 
This is illustrated in Figs.~\ref{fig_diffdyn2}, \ref{fig_diffdyn} and \ref{fig_diffdyn3}. 
In  Fig.~\ref{fig_diffdyn2} we show the observed and simulated PDFs. 
The observed PDF of this network time series is very similar to that of 
the mean which is shown in Fig.\ref{fig_differentalph}.  
Obviously, the chosen 
parameters of this particular simulation ($\alpha=2.24$,  $\tau_d+\tau_r=19.1$ time steps, 
$p_f=0.16$,   $y_{min}=0.014$,  $y_{max}=0.76$) %
 give an excellent fit to the observed PDF. 
A comparison with an actual time series (Fig.~\ref{fig_diffdyn}) shows, however, that the 
individual microflares last much longer in the simulations than in the data. Therefore, the 
probability distribution  cannot constrain the dynamics of the time series.
 In terms of the free parameters of the simulation, this implies that multiple combinations 
of free parameters can produce the same probability distribution, i.e., 
the same $\mu$ and $\sigma$ values. 
To demonstrate this 
we have varied $\alpha$ and $\frac{1}{\tau_d p_f}$ 
together, such that 
for smaller  $\alpha$ we have chosen a 
higher $\frac{1}{\tau_d p_f}$ in order to keep the %
$\sigma$ near the observed value. 
The observed values of the $\mu$ and $\sig$ parameters are generated 
using all the sets of simulation parameters given in Tab.~\ref{tab_alpha}.
\begin{table}
\caption{Values of $\alpha$, $p_f$, $\frac{1}{\tau_d p_f}, y_{min}$ and $y_{max}$, 
as well as the $\mu, \sigma$, and mean of the distributions produced by them. 
The values of the SUMER time series used for comparison (\oiv network) 
were $\mu_{SUM}= -2.16, \sigma_{SUM}= 0.98$, 
and mean$_{SUM}=0.15$.  $\Delta CDF$ gives the summed 
difference 
of the cumulative distribution functions.}  

\label{tab_alpha}
\begin{center} 
\begin{tabular}{cccccc} \hline
\small
$\alpha$ & 1.8  & 2.1 & 2.5 & 2.8 & 3.0 \\ 
\hline 
\hline
$\tau_d+\tau_r$  & 8.01 & 6.87 & 5.72 & 5.66 & 5.55 \\
$p_f$  & 0.60 & 0.52 & 0.43 & 0.43 & 0.42 \\  
$\frac{1}{\tau_d p_f} $  & 0.21 & 0.28 & 0.41 & 0.41 & 0.43 \\
$y_{min}$  & 0.002 & 0.006 & 0.016 & 0.019 &  0.021 \\
$y_{max}$  & 0.5 & 0.6 & 0.8 & 0.8& 0.8 \\
\hline
$\mu$  & $ -2.17 $ & $ -2.21 $ & $ -2.12$ & $ -2.12 $ & $ -2.16 $ \\
$\sigma$  & 0.95 & 0.97 & 1.0 & 0.97 & 0.95 \\
mean & 0.15 & 0.14 & 0.15 & 0.15 & 0.14 \\ 
$\Delta CDF$  & 0.007 & 0.008 & 0.004 & 0.007 & 0.013 \\
\hline
\end{tabular}
\end{center} 
\end{table}
The dynamics of the time series corresponding to the parameters given in 
Tab.~\ref{tab_alpha}, however, are in general not identical.
In particular, $\tau_d$ and $p_f$ influence the duration of the individual 
events and the interval between them. 
In order to constrain the time dependence, 
a diagnostic sensitive to time scales
of radiance variation is required. The (Fourier or wavelet) power spectrum 
is such a diagnostic. 
\begin{figure}[h]
\unitlength1cm
\centerline{\hbox{
\includegraphics[width=0.25\textwidth,clip=]{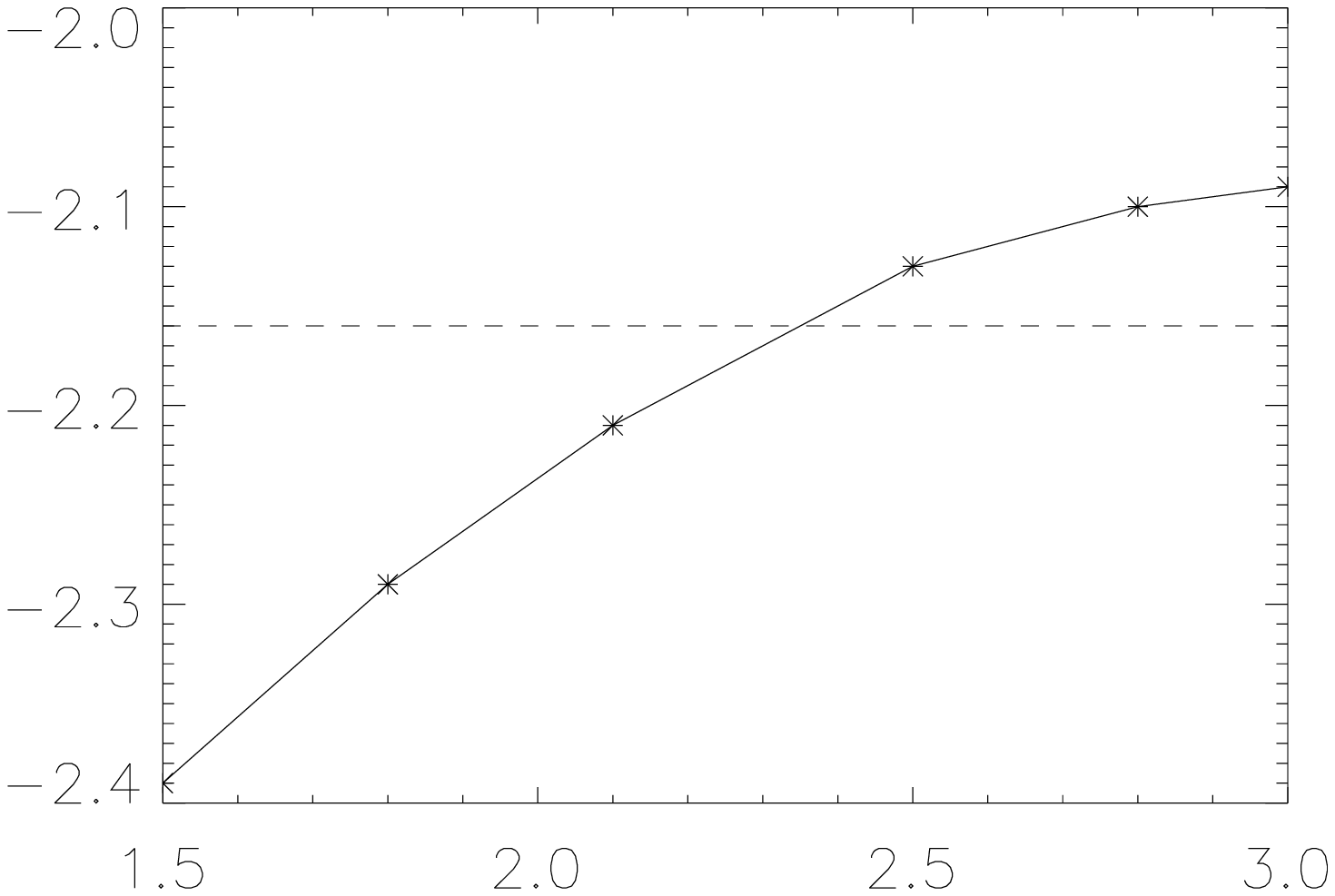}
\includegraphics[width=0.25\textwidth,clip=]{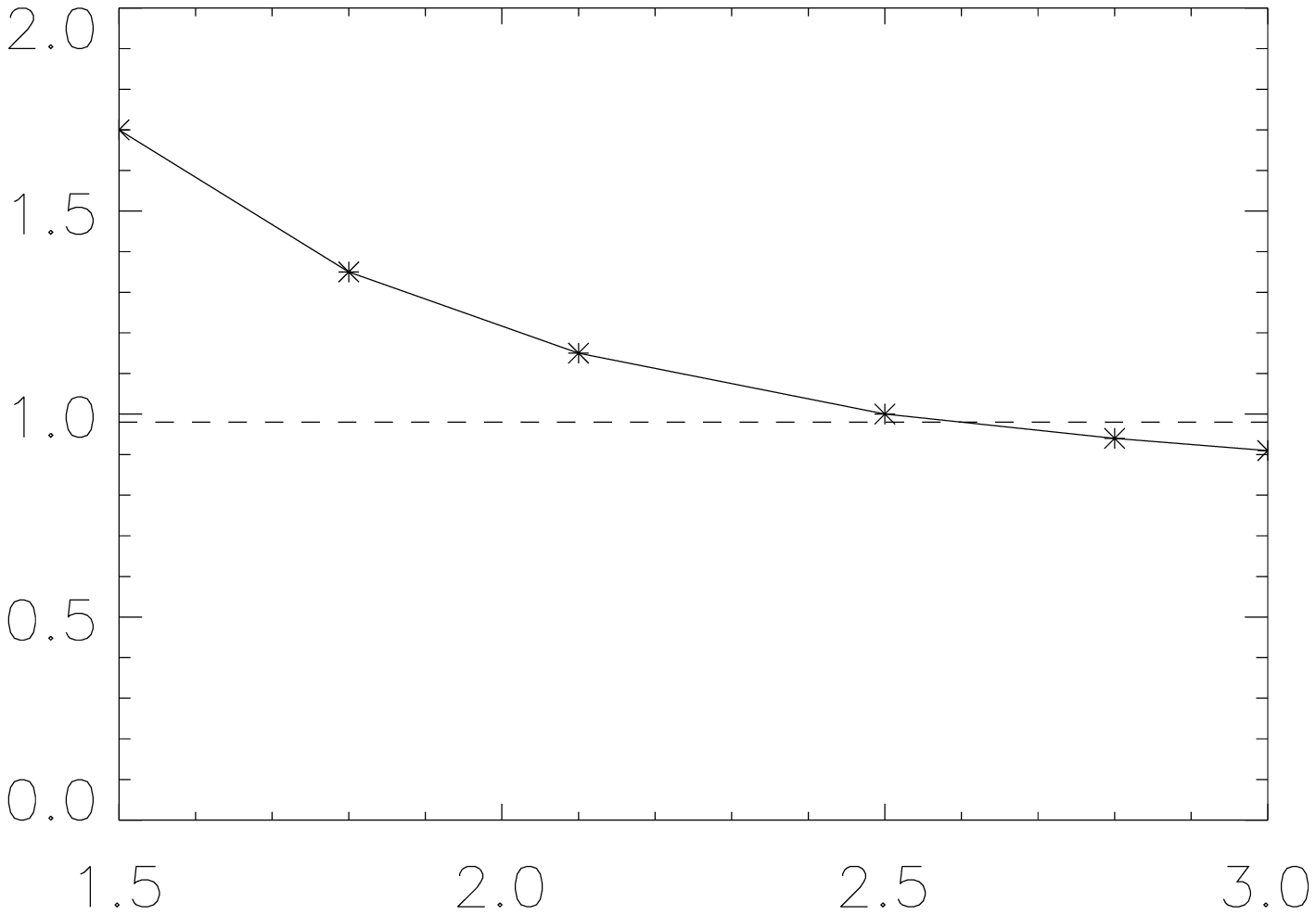}
}}
\begin{picture}(8,-.1)
 \put(0.0,2.2){$\mu$}  \put(4.5,2.2){$\sigma$}
\put(2.2,0.3){$\alpha$}  \put(6.6,0.3){$\alpha$}
\end{picture}
\caption[1]
{Variation of the lognormal  parameters with the exponent $\alpha$ of the 
flaring input distribution. The total energy input has been held constant by 
varying the lower boundary of the flare amplitudes accordingly. The dashed lines 
show the corresponding lognormal-fit values to the averaged SUMER \oiv 
network data.  
}
\label{fig_musigwithalpha}
\end{figure}
The power spectral density was 
computed here using (Morlet-)wavelets (see, e.g., Torrence \& Compo, 1998). 
This choice is driven by the smaller amount of fluctuations in this quantity 
than in Fourier power spectra, since wavelets introduce a smoothing in the 
spectral domain. 
In Fig.~\ref{fig_diffdyn3} we note that the main peak 
of the global wavelet power spectrum of the observed time series 
lies at nearly one half of the frequency of that of the model. Thus,
this diagnostic nicely complements the radiance distribution. 
As we show in Sect.~\ref{sumanalysis} the combination of 
distribution function and power spectrum leads to much stronger constraints 
on the free parameters of the model.  
We cannot rule out, however, that the distinct peaks seen in the power spectra are
partly due to
the restricted length of the time series and the thereby limited statistics.
Moreover, some properties of the flaring process are best seen by directly 
comparing with an observed time series, although  this comparison remains 
qualitative.

\begin{figure}[h]
\unitlength1cm
\centerline{\hbox{
\includegraphics[width=0.44\textwidth,clip=]{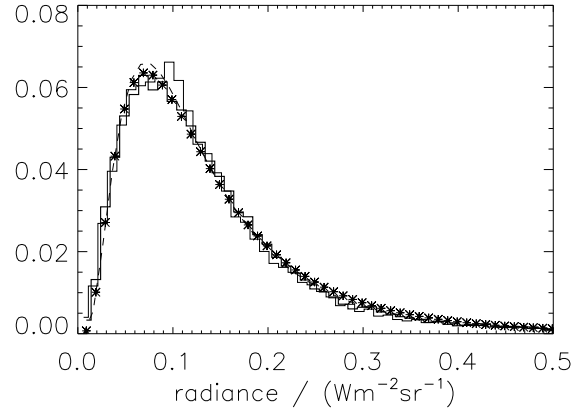} }}
\caption[1]
{The histograms corresponding to the time series of Fig.~\ref{fig_diffdyn}. 
The simulated histogram is represented by the thin solid line. 
The dashed line indicates the lognormal-based fit to this distribution. 
The SUMER histogram is given by the thick line and its fit is indicated 
by the stars. }
\label{fig_diffdyn2}
\end{figure}
\begin{figure}[h]
\unitlength1cm
\includegraphics[width=0.5\textwidth,clip=]{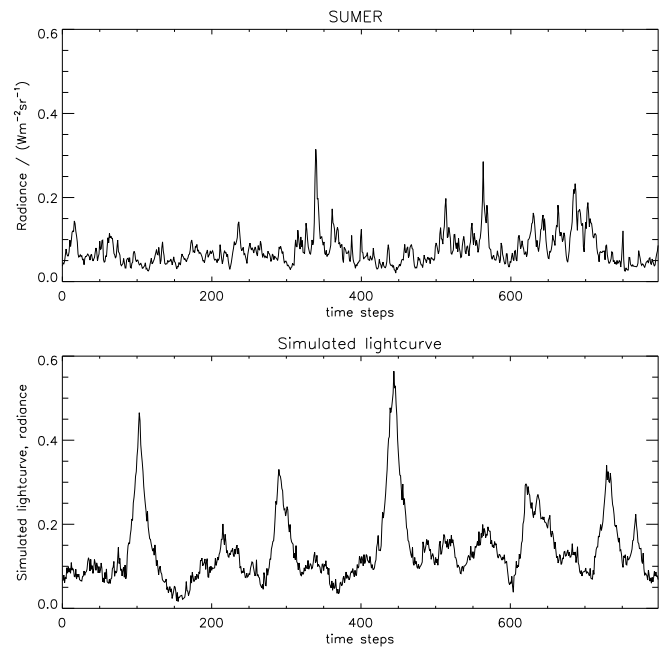}
\caption[1]
{Example of a simulation featuring a 
similar lognormal distribution as a SUMER (\oiv network) time series 
but a very different temporal variation,  %
so that the time series look rather different. 
Parameters of the simulation are $\alpha=2.24$,  $\tau_d+\tau_r=19.1$ 
time steps, %
$p_f=0.16$,   $y_{min}=0.014$,  $y_{max}=0.76$.
The corresponding histograms are shown in Fig.~\ref{fig_diffdyn2}.} 
\label{fig_diffdyn}
\end{figure}
\begin{figure}[h]
\unitlength1cm
\includegraphics[width=0.49\textwidth,clip=]{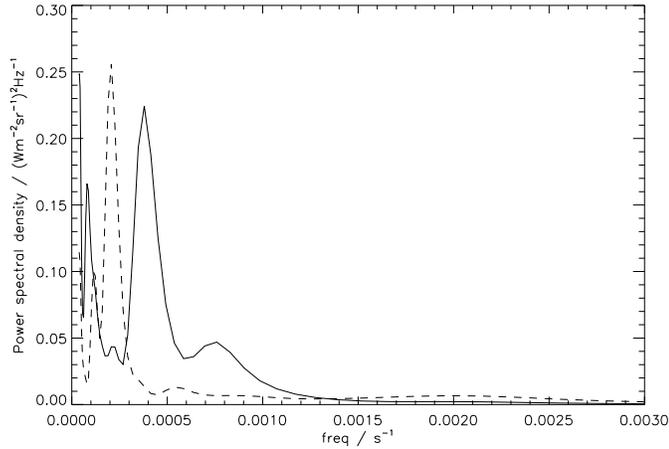}
\caption[1]
{The global wavelet power spectra of the 
modelled (solid line) and the observed (dashed) time series corresponding to Figs.~\ref{fig_diffdyn} and \ref{fig_diffdyn2}.}
\label{fig_diffdyn3}
\end{figure}

\section{Application of the model to SUMER transition region and coronal data}
\label{sumanalysis}

SUMER data of the EUV lines O\ts{$\scriptstyle {\rm IV}$},  \siv\ and 
Ne\ts{$\scriptstyle {\rm VIII}$} were used to 
constrain the free parameters of the model. 
These lines represent temperatures of the lower (cooler) transition region 
(O\ts{$\scriptstyle {\rm IV}$},  \siv) and the upper (hotter) transition 
region or lower 
corona (Ne\ts{$\scriptstyle {\rm VIII}$}). 
We restricted ourselves to areas where no significant change of the 
activity was discernible, i.e., truly quiet regions with no 
detectable trend in the average radiance during the measurements. 
The parameter values obtained from these data are listed in Tab.~\ref{tab_difftemp}. 
This table summarizes the optimum parameters for our model to represent the 
averages over the cell and network time series 
as suggested by comparison with the SUMER data. The summed difference of the cumulative 
distribution functions (compare Fig.~\ref{fig_differentalph}) is given in the last column.

The model parameters were varied until a good (statistical) match 
was obtained to the SUMER 
time series, its radiance distribution and its global wavelet power spectrum. 
We used a statistical power-law distributed flaring process and 
 introduced a possible rise time of a flare $\tau_r$, in order 
to better match the observed time series. 
The flare brightness was chosen to both, rise and fall exponentially. 
In all cases,  $\tau_r$ was chosen to be 75~\% of the damping time  $\tau_d$, 
so that the flaring time scale or ``effective damping time'' was  
$\tau_r+\tau_d$. 
Note that the need for a rise time $\tau_r$ comes neither from the radiance 
distribution functions nor from the power spectrum, 
but rather is indicated by the direct comparison between the time series 
(such as 
those shown in Figs.~\ref{fig_tr_cells} to~\ref{fig_corona_netw}).  
Each simulation consisted of $n=24\,000$ time steps. 
For the radiance histograms, a bin size of approximately 5~\% of the 
respective maximum value has been used. \\ 
We found that 
adding Poisson photon noise $(\sqrt{counts})$ to the model realizations 
gives slightly better matches when comparing the single time series 
(see Figs.~\ref{fig_tr_cells} and~\ref{fig_tr_netw}, and~\ref{fig_corona_cells}
 and~\ref{fig_corona_netw}) and their 
histograms.

\subsection{Transition region}

Figure~\ref{fig_histos} shows two sample histograms from %
one-pixel time series (i.e., from a 1\arcsec$\times$1\arcsec\ area on the 
solar surface)
of the SUMER  \oiv\ data, 
 the first from a darker area of the images, 
corresponding to the interior of a supergranule cell, 
the second from a brighter pixel (network). 
Also plotted in 
 Fig.~\ref{fig_histos} are the histograms %
resulting from  
simulations with the parameters given in the top two rows 
of Tab.~\ref{tab_difftemp}. 

The agreement between measured and modelled histograms is reasonable.  
The majority 
of the differences can be ascribed to the better statistics of the model.    
\begin{figure}[h]
\unitlength1cm
\includegraphics[width=0.48\textwidth,clip=]{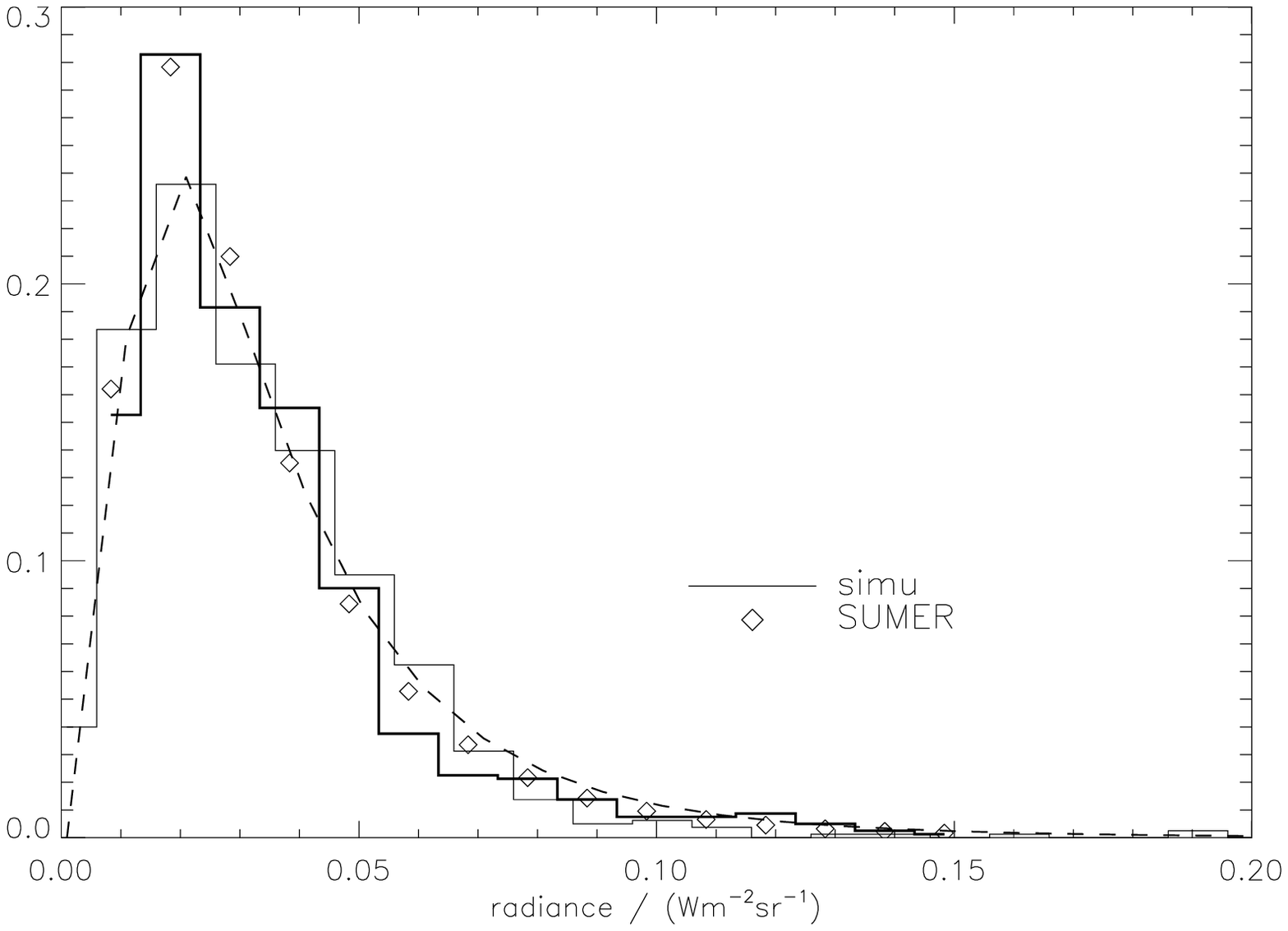}
\includegraphics[width=0.48\textwidth,clip=]{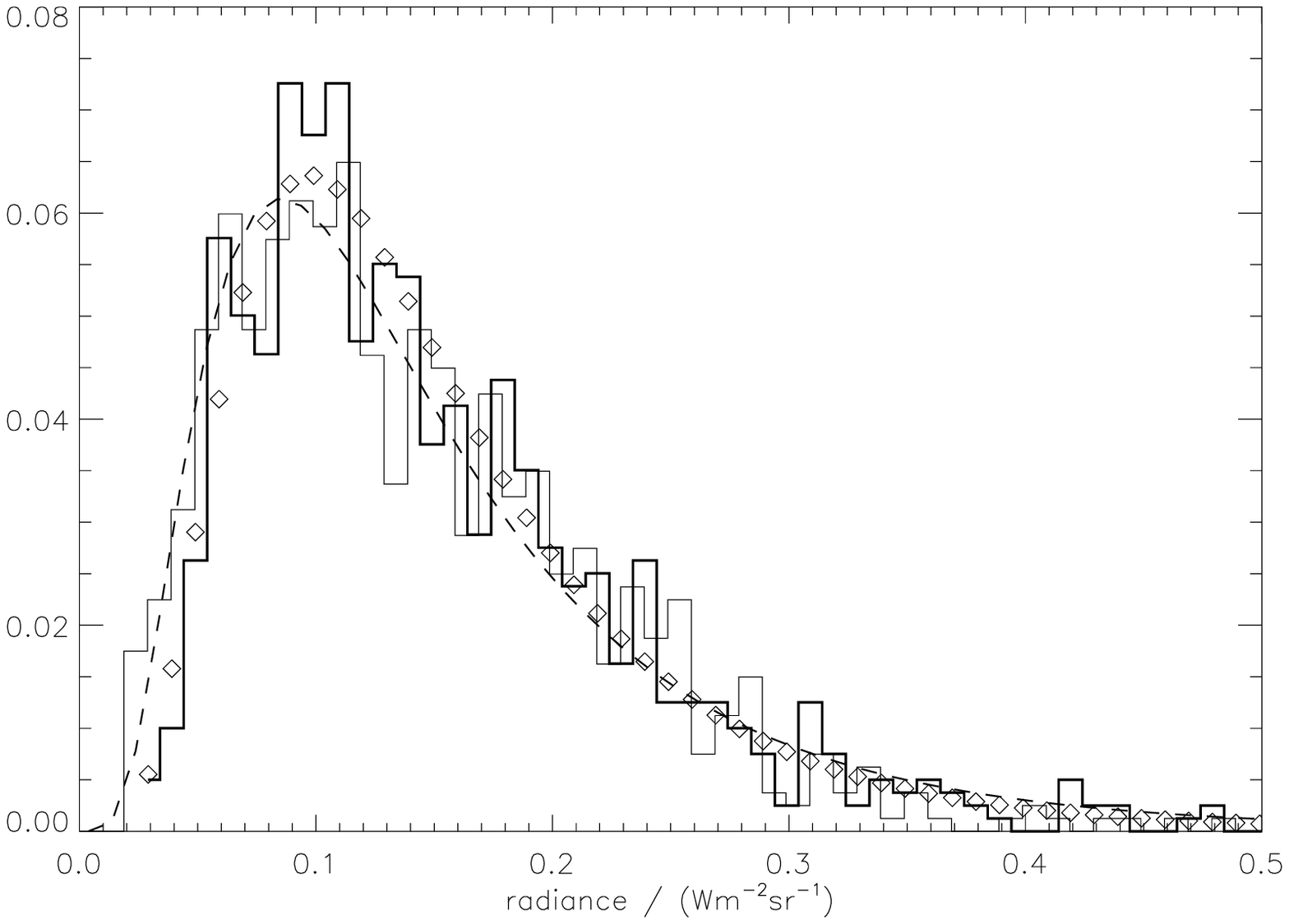}
\begin{picture}(8,-0.5)
\put(7.3,11.8){(a)}  
\put(7.3,5.6){(b)}  
\end{picture}
\caption[1]
{Histograms and lognormal fits of the simulated (thin-line histogram and  
dashed curves, respectively) 
and SUMER  (\oiv line at 79.0~nm)    %
measured (thick-line histogram and diamonds) radiances. 
(a): from a time series in a very quiet area (cell interior),  
(b): from a network sample of the same size (1\arcsec$\times$1\arcsec).}
\label{fig_histos}
\end{figure}

\label{sec_6}

\begin{figure} %
\unitlength1cm
\begin{center}
\includegraphics[width=0.5\textwidth,clip=]{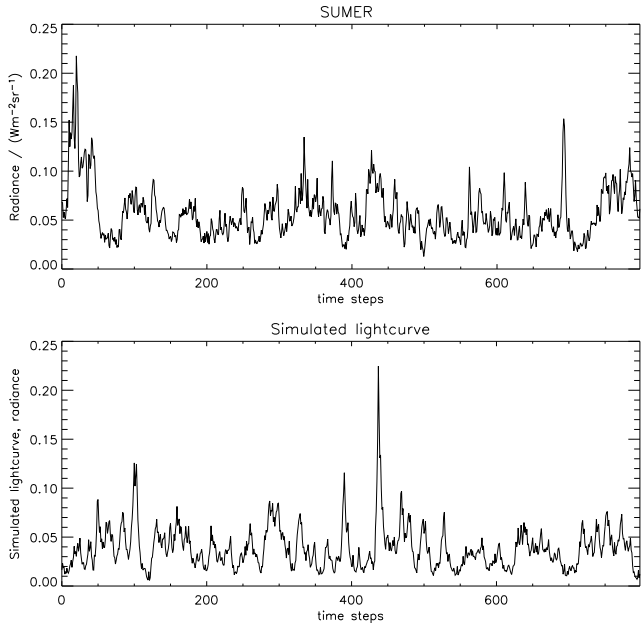}
\end{center}
\caption[1]
{SUMER time series of a darker (``cell'') region (1 pixel) in the 
transition region \oiv  79.0~nm line (upper frame),  
and a  corresponding simulation (lower frame).  }
\label{fig_tr_cells}
\end{figure}

Two examples of SUMER one-pixel time series of the \oiv line are shown in 
 Figs.~\ref{fig_tr_cells} and~\ref{fig_tr_netw}, together with 
the corresponding simulations. 
The photon noise  was on average 7~\% of the signal for the \oiv data. 
\begin{figure} %
\unitlength1cm
\begin{center}
\includegraphics[width=0.5\textwidth,clip=]{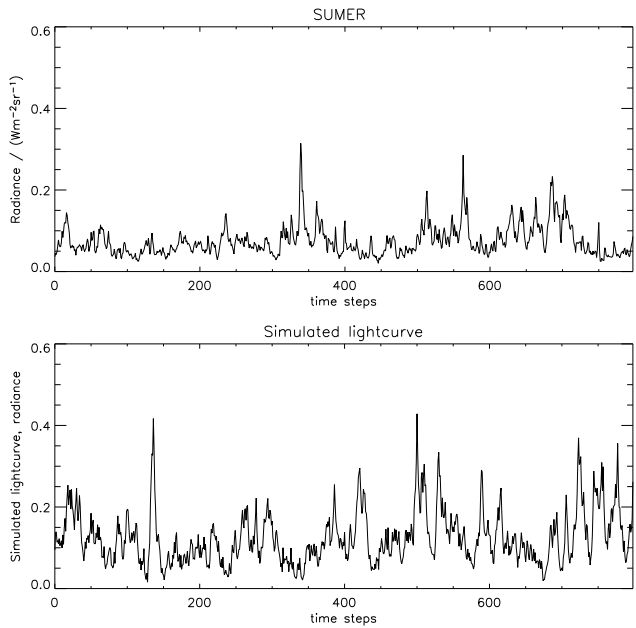}
\end{center}
\caption[1]
{SUMER time series of a brighter (``network'') region (1 pixel) 
in the transition region \oiv  79.0~nm line,  
 and a  corresponding simulation. } %
\label{fig_tr_netw}
\end{figure}

Figure~\ref{fig_psd790} gives the  measured and simulated 
power spectral densities, for both, the darker  and  the brighter region. 
The modelled and the observed power spectra now both display peaks at 
similar frequencies. 
It is noteworthy that 
the same model parameters %
reproduce both the network and the cell interior data. Only 
 the amplitude of the input process had to be adjusted by setting 
the energy of the smallest flares. 
The large $\alpha$ value of 2.5 is obtained because the model attempts  
to reproduce all the diagnostics simultaneously and is not restricted to reproducing 
the brightenings but also the background. 
\begin{figure} %
\unitlength1cm
\centerline{\hbox{
 \includegraphics[width=0.47\textwidth,clip=]{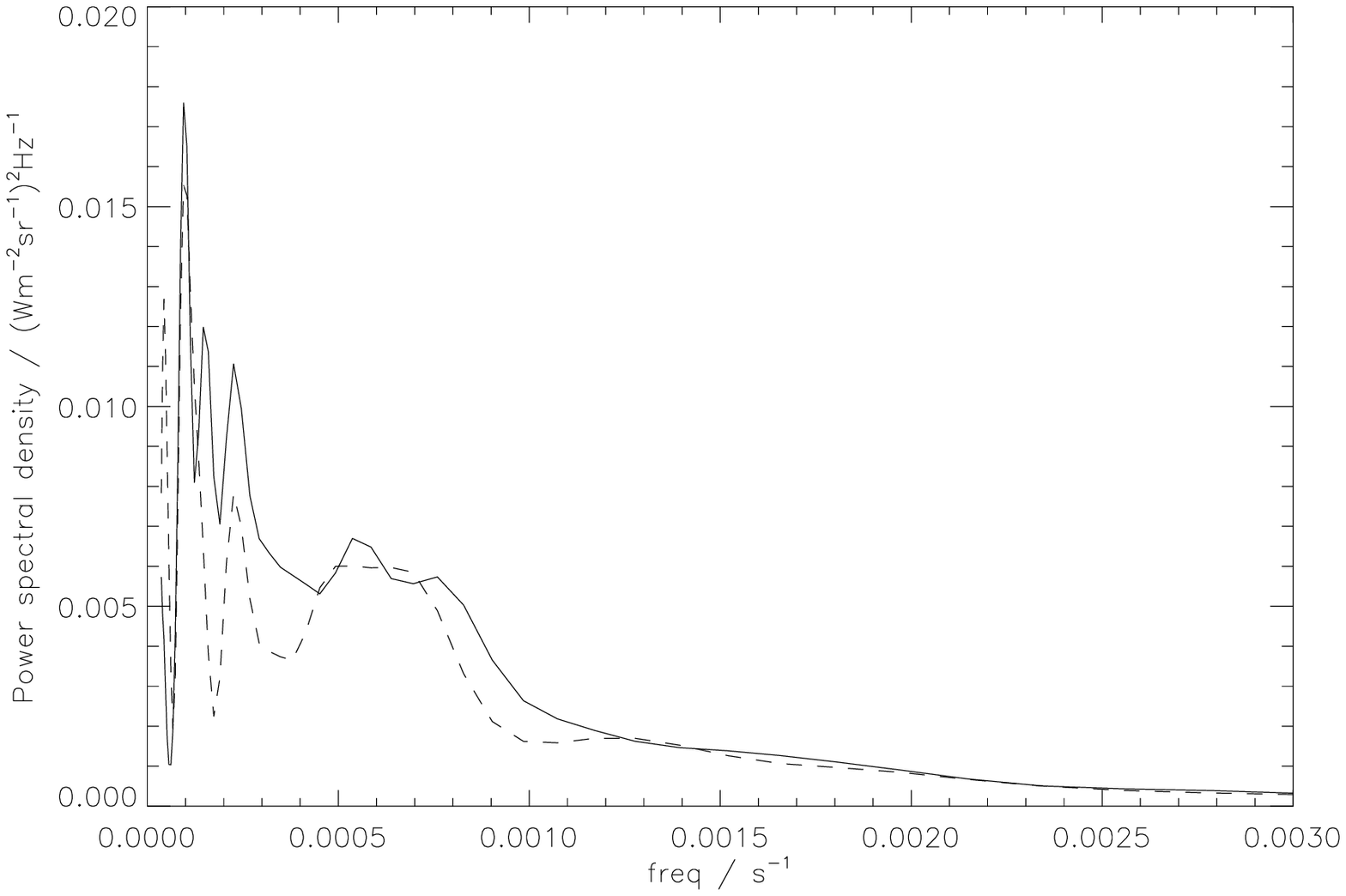}
 }}
\centerline{\hbox{
 \includegraphics[width=0.47\textwidth,clip=]{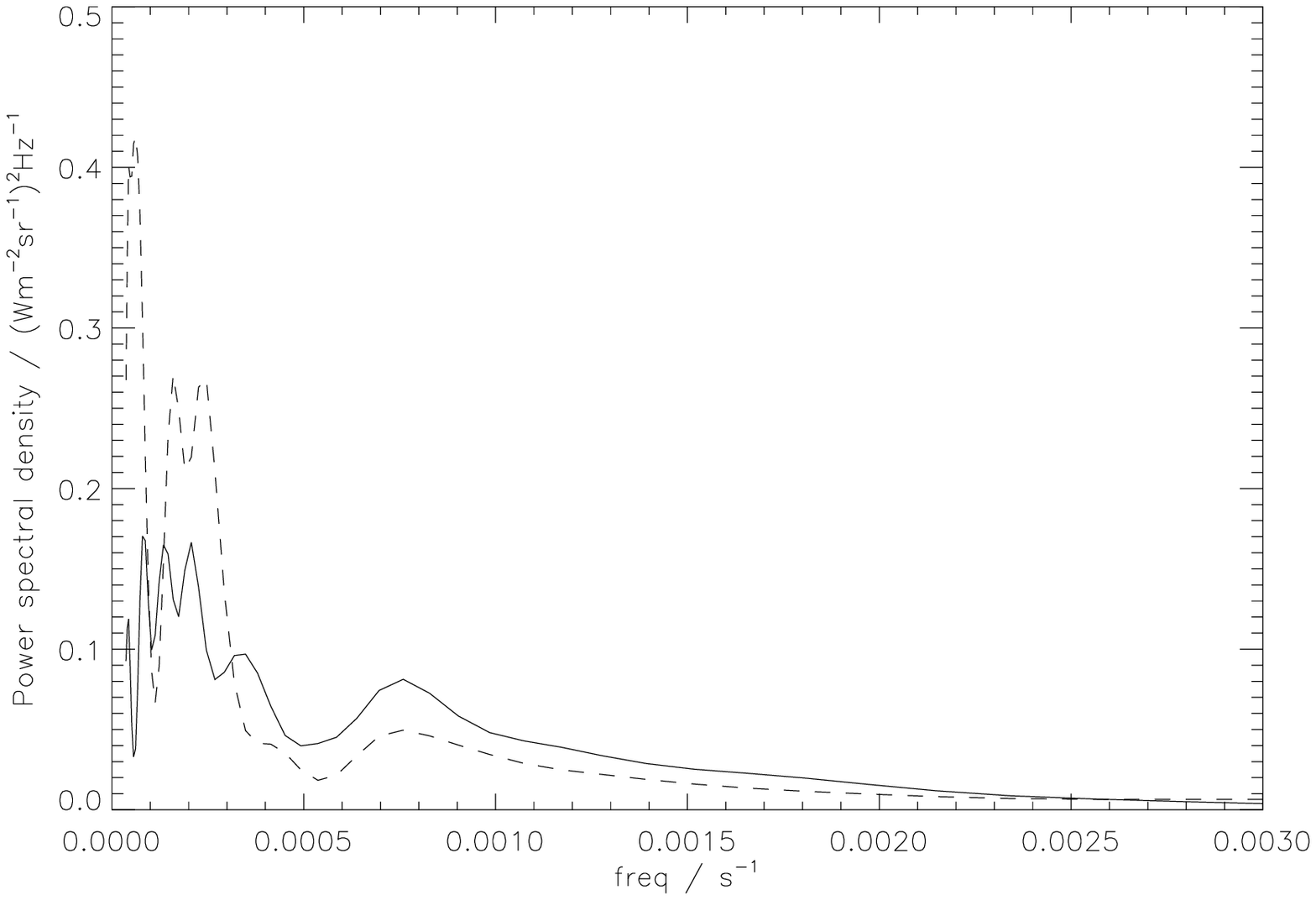}
}} 
\begin{picture}(5,0.)
\put(7.,11.){(a)}  \put(7.0,5.1){(b)}  
\end{picture}
\caption[1]
{Wavelet global 
power spectra for the time series shown in Fig.~\ref{fig_tr_cells} (a),   
and Fig.~\ref{fig_tr_netw} (b). The solid lines show the simulated
spectrum, and the dashed lines give the spectrum of the SUMER time series.}  
\label{fig_psd790}
\end{figure}

\subsection{Corona}

\begin{figure}[h]
\unitlength1cm
\includegraphics[width=0.48\textwidth,clip=]{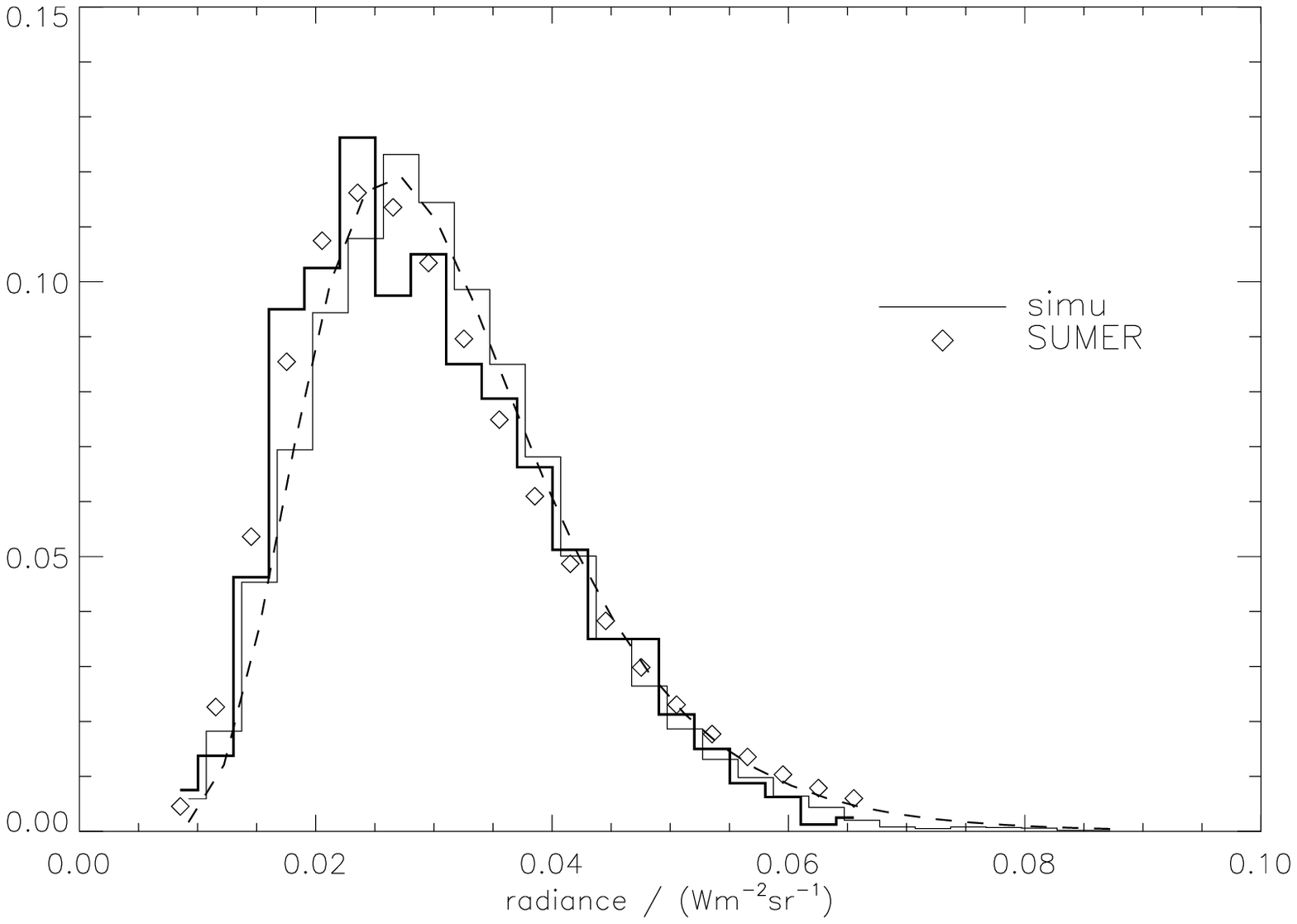}
\includegraphics[width=0.48\textwidth,clip=]{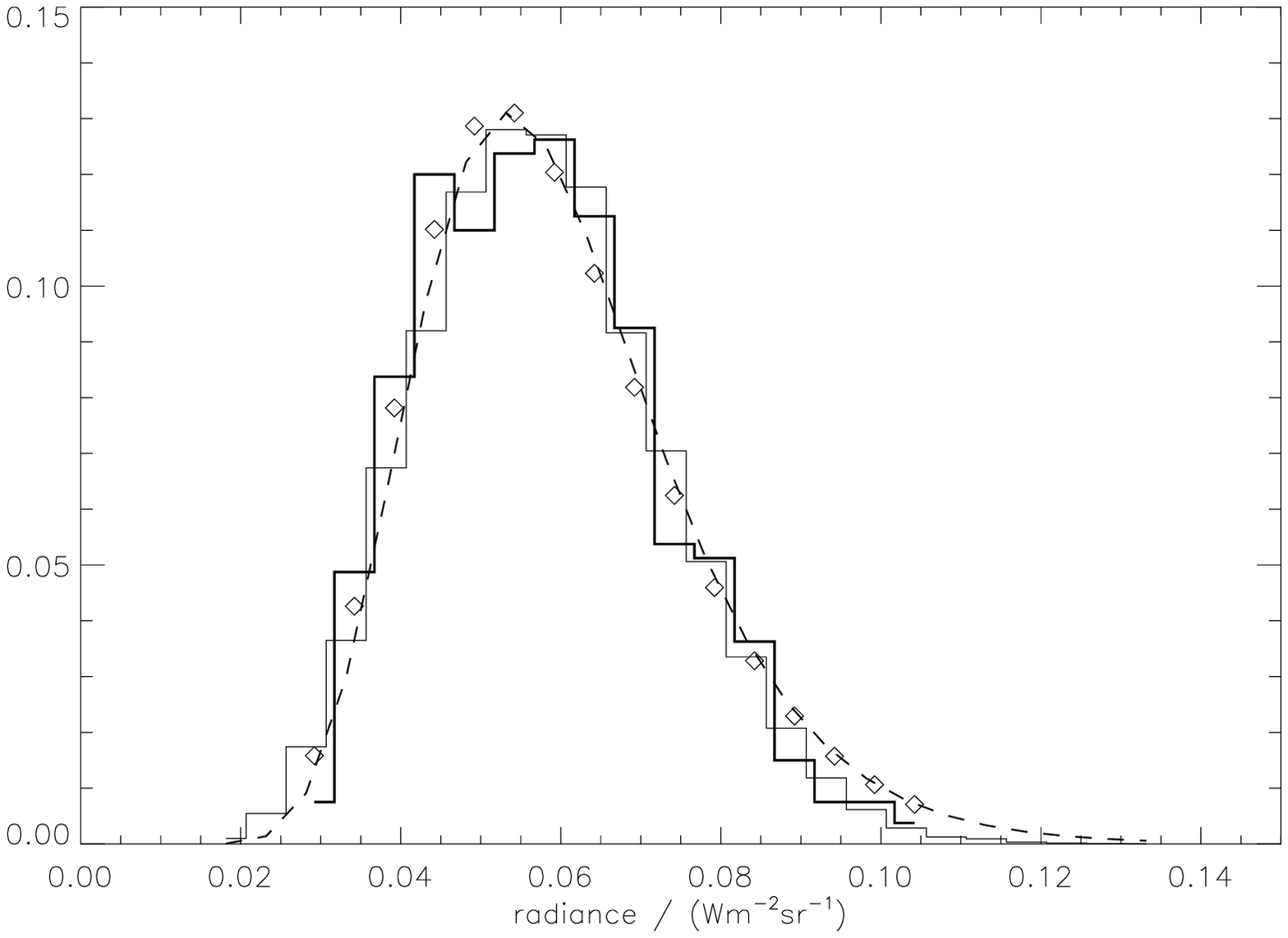}
\begin{picture}(8,-0.5)
\put(7.3,11.8){(a)}  
\put(7.3,5.6){(b)}  
\end{picture}
\caption[1]
{Histograms and lognormal fits of the simulated (thin-line histogram and 
dashed curves, respectively)  
and SUMER  (\ne8\ line at 77.0~nm)
measured (thick-line histogram and diamonds) radiances. 
(a): from a time series in a dark pixel (1\arcsec$\times$1\arcsec),  
(b): from a bright pixel.}
\label{fig_histos770}
\end{figure}

The coronal time series and the corresponding histograms as well as power spectra 
recorded in cell interiors could be well represented by a 
weakly damped and low-frequent flare excitation, 
with the parameters given in the  third row of Tab.~\ref{tab_difftemp}. 
The correspondence between model and data can be judged from Figs.~\ref{fig_histos770}a,
\ref{fig_corona_cells} and  \ref{fig_psd770}a. 
Roughly 9~\% photon noise was superposed on the modelled time series on average.  

\begin{figure} %
\unitlength1cm
\begin{center}
\includegraphics[width=0.5\textwidth,clip=]{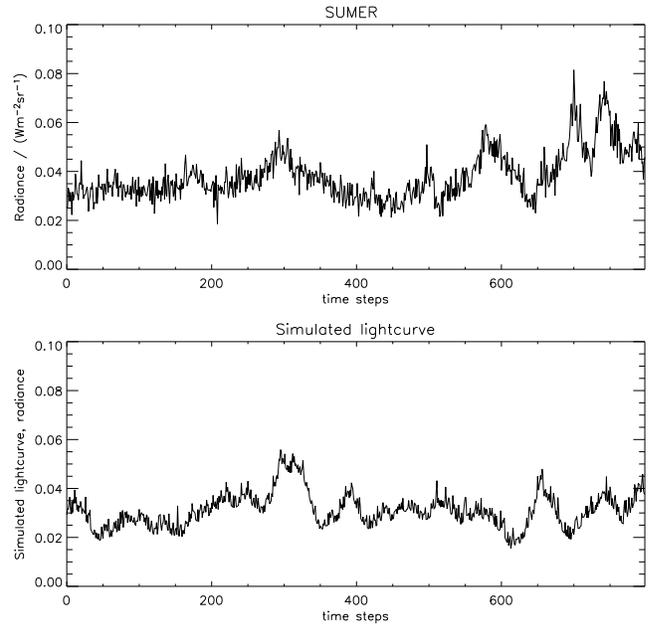}
\end{center}
\caption[1]
{SUMER time series of a darker region (1 pixel) in the 
upper transition 
region/coronal line 77.0~nm (\ne8), and a  corresponding simulation. } 
\label{fig_corona_cells}
\end{figure}
Satisfactory 
results for  brighter regions could also be obtained, as illustrated in 
Figs.~\ref{fig_histos770}b, 
\ref{fig_corona_netw} and  \ref{fig_psd770}b. 
The parameters are listed in the last line of  Tab.~\ref{tab_difftemp}. 
$\tau_d+\tau_r$ needed to describe the \ne8 time series and power spectra 
(Fig.~\ref{fig_psd770}) in the brighter parts of the quiet Sun are roughly 2.5 times 
larger than in the darker parts. The product $(\tau_d+\tau_r)/p_f$, however, differs by
a factor of only 1.5. The difference in this product is sufficient to explain the 
relatively small difference in 
the mean intensity (Tab.~\ref{tab_difftemp}), so that the same $y_{min}$ is used for bright 
and dark regions. 

\begin{figure} %
\unitlength1cm
\begin{center}
\includegraphics[width=0.5\textwidth,clip=]{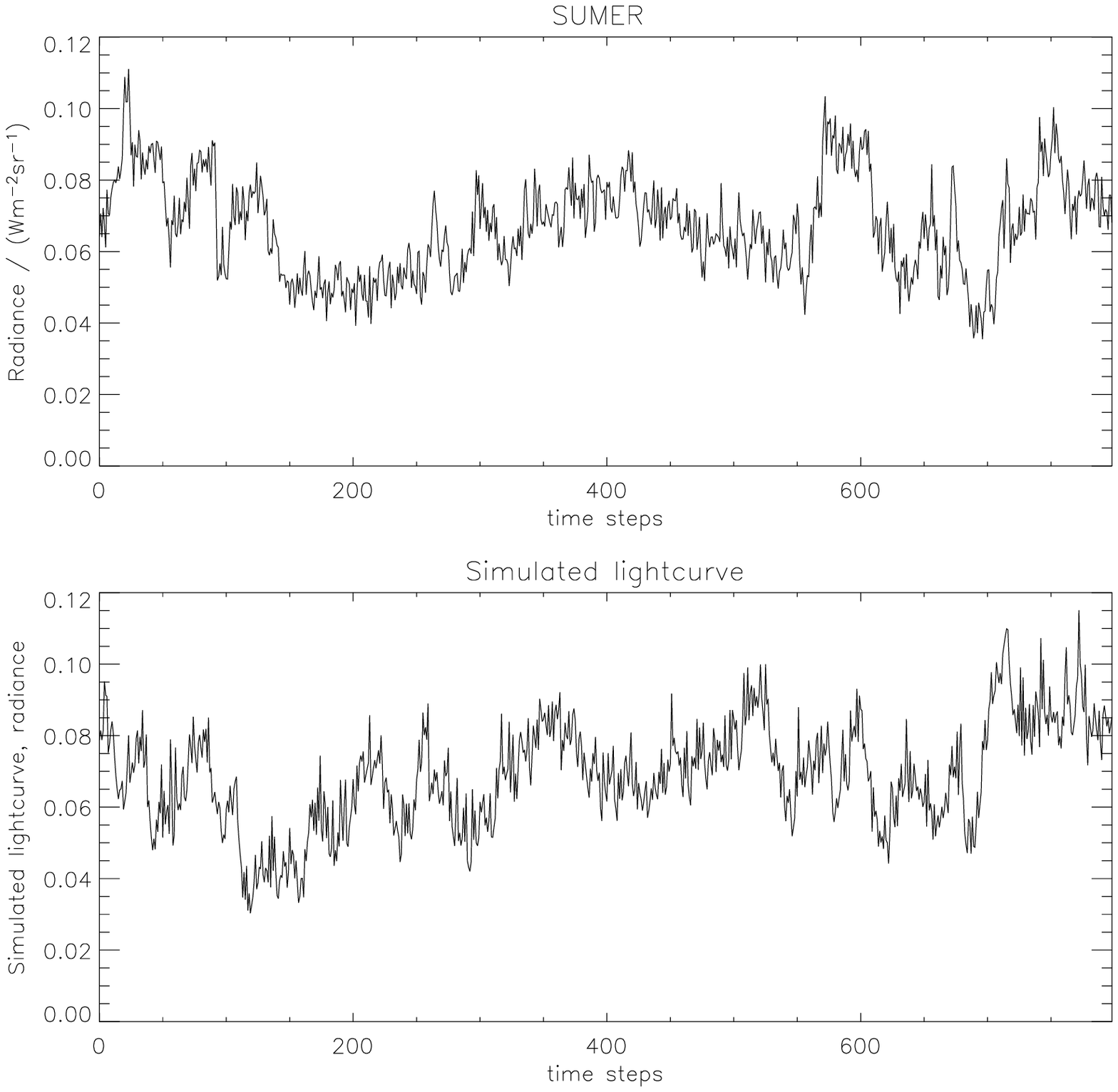}
\end{center}
\caption[1]
{SUMER time series of a brighter region (1 pixel) in the upper transition 
region/coronal line 77.0~nm (\ne8), and a  corresponding simulation.} 
\label{fig_corona_netw}
\end{figure}
\begin{figure} %
\unitlength1cm
\centerline{\hbox{
 \includegraphics[width=0.47\textwidth,clip=]{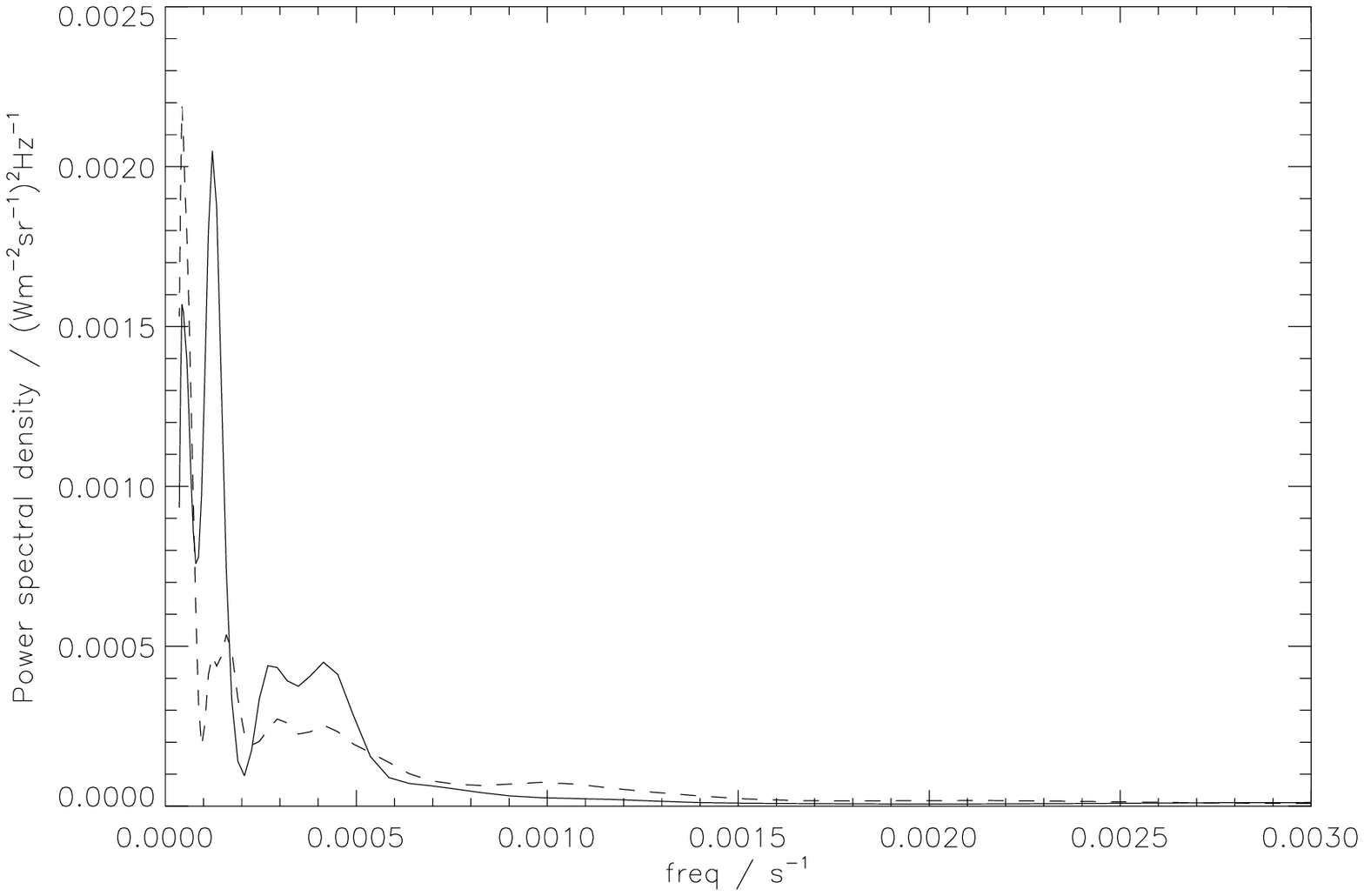}
 }}
\centerline{\hbox{
 \includegraphics[width=0.47\textwidth,clip=]{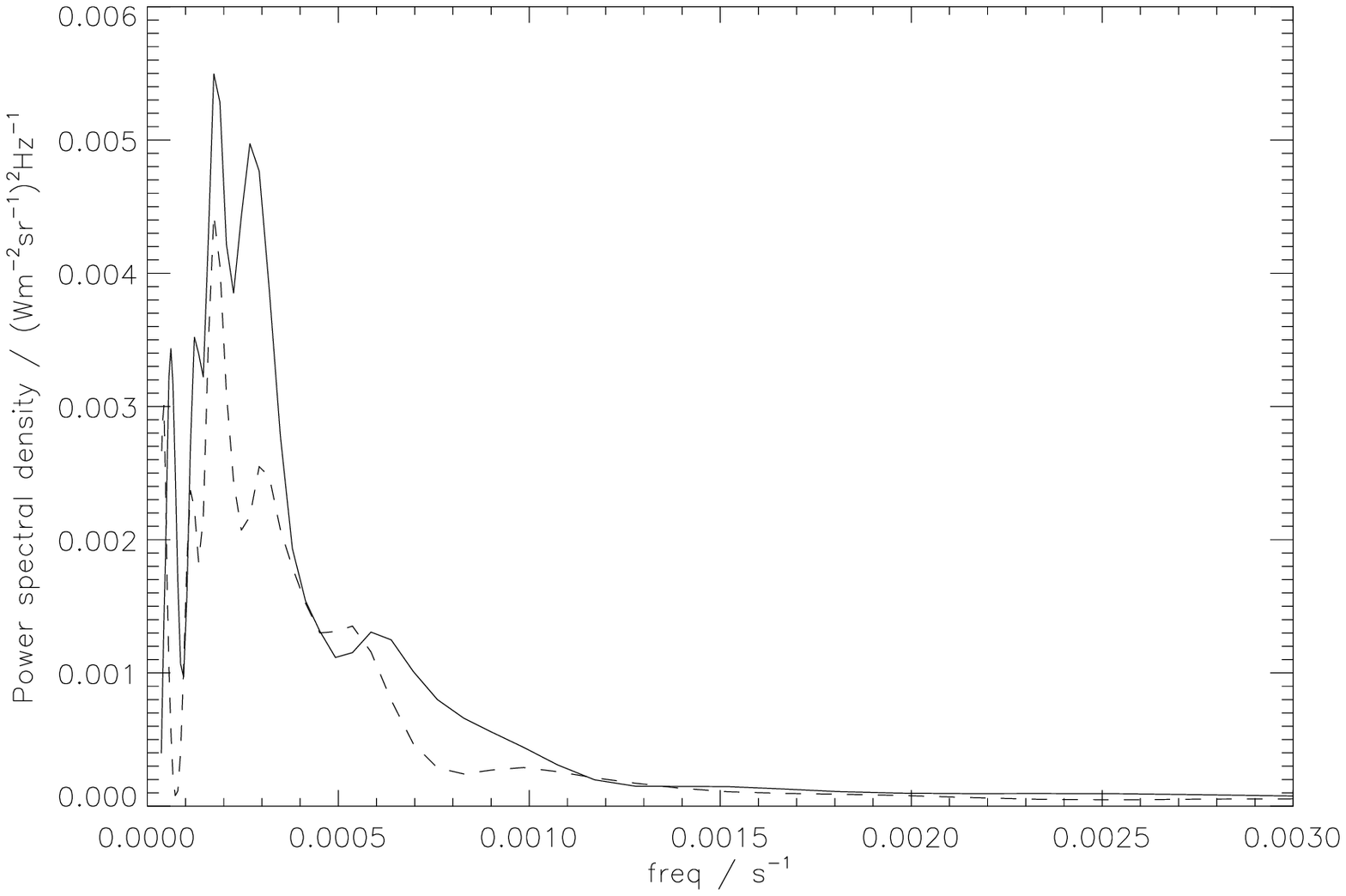}
}} 
\begin{picture}(5,0.)
\put(7.,11.){(a)}  \put(7.0,5.1){(b)}  
\end{picture}
\caption[1]
{Wavelet global 
power spectra for the time series shown in Fig.~\ref{fig_corona_cells} (a),  
and Fig.~\ref{fig_corona_netw} (b). The solid lines show the 
simulated
spectrum, and the dashed lines give the spectrum of the SUMER time series.}  
\label{fig_psd770}
\end{figure}
\begin{table*}
\caption{Best-fit parameters for the SUMER measurements 
of 8 February 1998.} 

\label{tab_difftemp}
\begin{center} 
\begin{tabular}{ccccccccc|ccc|c} \hline
\small
 & $\alpha$ & $\tau_d+\tau_r$  & $p_f$ & $y_{min}$ & $y_{max}$ & $\mu$ &
 $\sigma$ & mean & $\mu_{SUM}$ & $\sigma_{SUM}$ & mean$_{SUM}$ & 
$\Delta CDF$ \\
\hline
\hline
dark px. 79.0 nm& 2.5 & 5.721 (191.68 s) & 0.43 & 0.004 & 0.80 & $-3.54$ & 
1.08 & 0.042 & $-3.58$ & 1.09 & 0.039 & 0.0045 \\ 
bright px. 79.0 nm & 2.5 & 5.721 (191.68 s) & 0.43 & 0.016 & 0.80 & $-2.12$ & 1.0 & 0.15 & $-2.16$ & 0.98 &  0.15 & 0.0043 \\ 
dark px. 77.0 nm &2.5 & 44.40 (1487.5 s) & 0.11 & 0.003 & 0.01 & $-3.67$ & 0.47 &
 0.027 & $-3.67$ & 0.45 & 0.028 & 0.0005 \\
bright px. 77.0 nm &2.5 & 18.28 (612.5 s) & 0.43 & 0.003 & 0.01 & $-2.86$ & 0.39  & 0.06 & $-2.84$ & 0.39 & 0.06 & 0.0022 \\
\hline
\end{tabular}
\end{center} 
\end{table*}
\subsection{Physical parameters}

Clearly, our model is too simple to give, on its own, 
 an accurate energy range of the 
flaring events that reproduce the data. However, with some restrictive 
assumptions a limited estimate can be made using Eq.~(\ref{eqenergy}).

The amplitudes listed in Tab.~\ref{tab_difftemp}
correspond to flare energies 
 between  
$6.4\times10^{16}$~erg %
and $1.3\times10^{19}$~erg  %
for \oiv and between  
$4.8\times10^{16}$~erg  %
and $1.6\times10^{17}$~erg %
for \ne8 for 
events of 1~s damping time covering an area of 1 square arcsecond  
(i.e., energy fluxes in the range between 12.6 and 2520~erg\,s$^{-1}$cm$^{-2}$
for \oiv and between 9.4 and 31.5~erg\,s$^{-1}$cm$^{-2}$ for 
Ne\ts{$\scriptstyle {\rm VIII}$}).

One grossly simplifying assumption that has already entered Eq.~(\ref{eqenergy}) 
concerns the geometry (radiation from a flat surface with no centre-to-limb 
variation). Another is that all events cover the same solar surface area, $A$. 
A third is that all events have the same temperature (or rather the same 
temperature-density relationship), so that a single $q$ value can be employed for all. 
We do not attempt to compute $q$, so that we only estimate the energy radiated 
in the observed spectral line. For the area $A$ 
we assume a size of $10^{13}$m$^2$, which 
would roughly be 20 SUMER pixels  
(\cite{Aschw00} found 
sizes of  $4 - 200\times10^{12}$m$^2$ using TRACE data),  
and also adopt durations of 200 to 2000 seconds.  %
 Under these assumptions we obtain a total range of energies between 
$2.52\times10^{20}$~erg to $5\times10^{24}$~erg for 
 individual events in the \oiv line and 
$1.88\times10^{20}$~erg to $6.3\times10^{21}$~erg for events in the \ne8 
line. 
Of course, this energy radiated away by only single spectral lines is 
still too low by several  
 orders of magnitude to make up for the upper chromospheric or coronal 
energy losses given, e.g., by \cite{withbroenoyes77}. 

From the flaring frequencies given in Tab.~\ref{tab_difftemp}, again  
assuming an event size of 20 SUMER pixels, one can obtain a rough estimate 
of  the number of such events on the total solar surface. 
To reproduce the SUMER time series in the transition region line and bright
corona including the background, it would need roughly 7600 events per second. 
2000 events per second can generate the radiation in the darker areas of 
the \ne8 line.
Given all the uncertainties entering the above estimates we prefer to restrict 
ourselves to order of magnitude estimates: our model suggests a nano-flaring rate 
of  $10^{3}$ to $10^{4}$ events per second. 
 These values can be compared with the number of blinkers 
estimated for the total solar surface (50 events/s, Brkovi{\' c} et al., 2000) and 
that of explosive events  (500 events/s, Innes et al., 1997). 

To estimate the total available flare input for heating 
one would need to know or extrapolate these values to the entire spectrum, 
assuming as precise areas as possible (geometry, filling factor) and 
prescribe which fraction of the total released energy is radiated away 
(and not, e.g., employed to accelerate the solar wind). 
Here we take a simpler approach. 
In order to  get an idea of the energy released by a typical brightening, 
we use 
 the total coronal radiative energy loss of the quiet Sun 
$F_r = 10^{5}$~erg cm$^{-2}$ s$^{-1}$ (corresponding to 
$6.07861\times10^{27}$~erg s$^{-1}$ over the entire solar surface area)
given by \cite{withbroenoyes77}. 
Assuming 7600 events per second, we get 
a mean radiative loss of $8\times10^{23}$~erg  per event  
which is an integral over the full spectrum and an average over the darker and brighter 
network areas. 

If we further assume the same power-law exponent $\al=2.5$
over the entire spectrum, %
we can estimate the lower energy limits of the 
flaring events obtained from our analysis from Eq.~\ref{eq_flareenergy}, 
\begin{eqnarray}
 m_{all} = 3 \frac{(y_{max}^{-0.5} - y_{min}^{-0.5})}{(y_{max}^{-1.5} - y_{min}^{-1.5})} 
         \longrightarrow  y_{min} \approx 8\times10^{23}~{\rm erg}  %
\label{eq_allflaremin}
\end{eqnarray}

for 7600 events per second. Here we have further assumed that $y_{max} >> y_{min}$.  
For 2000 events per second (as was estimated for the darker parts of the corona with smaller 
flaring frequency and much larger damping time), the average flare energy would approximate 
$3.04\times10^{24}$~erg. 
These values would make the brightening events in our analysis nano- and microflares, with 
the smallest ones being rather small nanoflares or large picoflares, depending on where the 
lower energy limit for flares is set. 
\cite{parker88} estimated values of around $1.6\times10^{24}$~erg 
per nanoflare event.

\section{Discussion and Conclusions}

In this paper we consider the diagnostic content of radiance time series 
and simulate them with the help of a simple stochastic model. 
The radiances of time series measured by SUMER in the quiet Sun 
are distributed following a 
lognormal function for every spatial point. The properties of these 
distributions reflect those found by \cite{pauluhn00} in the sense that 
transition region radiation shows broader, more asymmetric distributions, 
while the coronal radiance exhibits a narrower, more symmetric distribution. 
These results show that quiet-Sun radiances exhibit a lognormal distribution 
irrespective of whether we sample spatially \citep{pauluhn00} or temporally 
(this paper), although in the former case we are mixing radiation from 
network and supergranule cells, while in the latter case we are considering 
these features separately.

The shapes of the lognormals (Figs.~\ref{fig_histos} and \ref{fig_histos770}) 
emphasize the lack of 
points around a radiance of zero, with lines at higher temperatures showing 
this effect more strongly. 
In the past such a behaviour has often been 
interpreted to indicate the superposition of a more or less steady 
background and overlying peaks produced by discrete energy releases, e.g., 
by microflares. 
Here we assume that all the radiance in the observed lines is caused by 
micro- and nanoflares, so that any background is due to a superposition 
of many small brightenings.  
 
We have developed a model which attempts to explain the properties 
of the observed time series assuming that all the radiation coming from 
the transition region and the corona is due to the cooling of gas that 
had been heated by nanoflares. 
From the multiplicative version of the central limit theorem it follows that 
the product of many independent, identically distributed, positive random 
variables has approximately a lognormal distribution. (For a discussion 
of the special features of the lognormal distribution function see 
also, e.g., Limpert et al., 2001.)  
For a lognormal distribution to form in a simulated time series, the input 
process has 
to be positive definite (thus, 
a Gaussian with zero mean as input distribution does not produce a 
lognormal). 
By the positive input process, the asymmetric shape is ensured.

The shape and scale of the resulting lognormal are strongly dependent  
on the ratio of the damping to the excitation process. 
The higher the flaring frequency and the longer the damping time, the 
more symmetric the resulting lognormal becomes (i.e., the $\si$ values 
which determine the shape of the lognormal become smaller and the 
distribution  
becomes more Gaussian-like) and vice versa. 
The scaling of the distribution, which is its extension in (peak-)  height 
and width and is given by the parameter $\mu$ of the lognormal, 
increases with damping time 
(i.e., with decreasing damping) and with flaring frequency. 
Consequently, if many small brightenings overlap, they produce a stronger 
``background'' emission than fewer brightenings lying further apart. 
It has to be noted that the SUMER time series could be reproduced in a 
model employing the same damping time for weak and strong flares. 

An important part of this work has been to identify representations 
of the data which can serve as diagnostics to constrain a microflare model  
and to determine the free parameters, in particular the 
power-law exponent $\alpha$ of the flare amplitudes or energies. 
The combination of the 
power spectrum of the radiance time series and 
the  distribution function, %
together with the time series itself are found to constrain the free 
parameters of the model, including $\alpha$, relatively well. 
The most important result of this paper is that 
simulations  that best reproduce the data 
had  power-law exponents $\alpha$ of around 2.5. 
We stress that we have not carried out a thorough uniqueness 
analysis here, although numerous attempts to reproduce the data with 
$\alpha < 2$ were unsuccessful. 
Note also that this value is obtained assuming that all (nano-)flares 
have the same geometry and cover the same area. The large $\alpha$  
is determined by the need to reproduce the background emission  
seen in the transition region and the corona. Smaller  $\alpha$ values 
give too much weight to single larger flares which do not reproduce 
the purely quiet time series (cf., Fig.~\ref{fig_diffdyn}). 
By assuming that all micro- and nanoflares cover the same surface area of 
$10^{13}$m$^2$ following \cite{Aschw00}, we find a nanoflaring rate of 
$10^{3} - 10^{4} $s$^{-1}$.

An interesting result of the current investigations is that the 
damping time scales of the simulations reproducing the coronal time series 
are found to be significantly larger 
than in the transition region (by a factor of 4 to 8).
A simple estimate 
of the radiative cooling times based on the Rosner formula gives
(cf., Rosner et al. 1978): 
\vspace{-0.1cm}
\begin{equation}
\tau_{rad} = 3 \frac{1.38\times 10^{-16} }{n_e \xi T^{a -1}}. 
\end{equation}
\vspace{-0.1cm}
For the temperature of formation of the \oiv line, $T=1.5\times  10^{5}$~K, 
$\xi=6.3\times 10^{-22}$, $ a=0$, while for  $T=6\times 10^{5}$~K, appropriate 
for Ne\ts{$\scriptstyle {\rm VIII}$}, 
$\xi=3.98\times 10^{-11}$, $a=-2$. If we assume a constant pressure of 
0.3 dyn cm$^{-2}$ (= 0.03 N m$^{-2}$), we obtain a ratio of 
$\tau_{rad}$(Ne\ts{$\scriptstyle {\rm VIII}$})/$\tau_{rad}$(O\ts{$\scriptstyle {\rm IV}$}) $\approx$ 100, which is 
in the same direction, as that  obtained from the analysis, 
although an order of magnitude larger. 

The similar flaring probability, $p_f$, obtained for (the brighter part of the quiet)  
corona and transition region does not allow us to 
determine whether 
the brightnings in the corona and the transition region are due to the same or 
to different flaring events based on this analysis.

Further work is needed, both to demonstrate the uniqueness of the deduced 
parameters, in particular $\al$, to increase the amount of realism in the 
model, and to consider more data. 
The first may be achieved by automating the process of finding 
a best statistical fit to the data (i.e., best simultaneous fit to 
the distribution 
function and the power spectrum), possibly coupled to a Monte Carlo simulation 
or involving genetic algorithms. 
An analysis of lines sampling higher temperatures is required in order to 
reveal the behaviour of hot coronal gas. 
There is also a need to statistically simulate time series of images.  
Such data may also provide some information on the size distribution 
of flares.    
Since a single time series belonging to a particular pixel may sample only 
some spatially limited parts of particular brightenings, the parameters 
obtained by statistically reproducing it with a 1-d model need not be 
entirely representative of the real Sun. Thus, higher-dimensional models 
could be of great advantage. 

Finally, an application of the analysis developed here to active regions 
could reveal similarities and differences between the flaring processes 
in active regions and the quiet Sun. 

\begin{acknowledgements}
SOHO is a project of international cooperation between ESA and NASA. 
The SUMER project is financially supported
by Deutsches Zentrum f\"ur Luft- und Raumfahrt (DLR), the Centre National
d'Etudes Spatiales (CNES), the National Aeronautics and Space
Administration (NASA), and the ESA PRODEX programme (Swiss contribution). 
Wavelet software was provided by C. Torrence and G. Compo, and is available 
at URL: \url{http://paos.colorado.edu/research/wavelets/}. 
We are grateful to 
Davina Innes, Luca Teriaca and an anonymous referee 
who made valuable suggestions and helped 
substantially improve the manuscript. 
\end{acknowledgements}
\vspace{-0.4cm}

\end{document}